\documentclass[fleqn,usenatbib,useAMS]{mnras}

\usepackage{amsfonts}
\usepackage{amssymb}
\usepackage{hyperref}
\usepackage{color}
\usepackage{float}
\usepackage{subfigure}
\usepackage{tabularx}
\usepackage{soul}
\usepackage{graphicx}	% Including figure files
\usepackage{amsmath}	% Advanced maths commands
\usepackage{amssymb}	% Extra maths symbols
\usepackage{multicol}        % Multi-column entries in tables
\usepackage{bm}		% Bold maths symbols, including upright Greek
\usepackage{pdflscape}	% Landscape pages

\usepackage[T1]{fontenc}
\usepackage{ae,aecompl}

\usepackage{newtxtext,newtxmath}
\title{Kaniadakis holographic dark energy: observational constraints and global 
dynamics}

\author[A.  Hern\'andez-Almada, et. al.]{A.  Hern\'andez-Almada$^{1}$
\thanks{Contact e-mail: \href{mailto:ahalmada@uaq.mx}{ahalmada@uaq.mx}},  Genly Leon$^2$,  Juan Maga\~na$^3$,  Miguel A. Garc\'ia-Aspeitia$^{4,5,6}$,  
\newauthor V. Motta$^7$, Emmanuel N. Saridakis$^{8,9,10}$,  Kuralay Yesmakhanova$^{10,11}$%
\\
$^{1}$Facultad de Ingenier\'ia, Universidad Aut\'onoma de 
Quer\'etaro, Centro Universitario Cerro de las Campanas, 76010, Santiago de 
Quer\'etaro, M\'exico,\\
$^2$Departamento  de  Matem\'aticas,  Universidad Cat\'olica del Norte, Avda.   Angamos  0610,  Casilla  1280  Antofagasta,  Chile,\\
$^3$Instituto de Astrof\'isica \& Centro de Astro-Ingenier\'ia, Pontificia Universidad Cat\'olica de Chile, Av. Vicu\~na Mackenna, 4860, Santiago, Chile,\\
$^4$Depto. de Física y Matemáticas, Universidad Iberoamericana Ciudad de México, Prolongación Paseo de la Reforma 880, México D. F. 01219, México\\
$^5$Universidad Aut\'onoma de Zacatecas, Calzada Solidaridad esquina con Paseo a la Bufa S/N C.P. 98060, Zacatecas, M\'exico,\\
$^6$Consejo Nacional de Ciencia y Tecnolog\'ia, Av. Insurgentes Sur 1582. Colonia Cr\'edito Constructor, Del. Benito Ju\'arez \\ C.P. 03940, Ciudad de M\'exico, M\'exico,\\
$^7$Instituto de F\'isica y Astronom\'ia, Facultad de Ciencias,  Universidad de Valpara\'iso, Avda. Gran Breta\~na 1111, Valpara\'iso, Chile,\\
$^8$National Observatory of Athens, Lofos Nymfon, 11852 Athens,  Greece,\\
$^9$CAS Key Laboratory for Researches in Galaxies and Cosmology,  Department of Astronomy, \\ University of Science and Technology of China, Hefei, Anhui 230026, P.R. China,\\
$^{10}$Ratbay Myrzakulov Eurasian International Centre for Theoretical Physics, Nur-Sultan 010009,  Kazakhstan,\\
$^{111}$Eurasian National University, Nur-Sultan Astana 010008, Kazakhstan.
}

\date{Last updated 2020 June 10; in original form 2013 September 5}

\pubyear{2021}

\begin{document}
\label{firstpage}
\pagerange{\pageref{firstpage}--\pageref{lastpage}}
\maketitle

\begin{abstract}
We investigate Kaniadakis-holographic dark energy   by confronting it with  observations. We perform a Markov Chain Monte Carlo analysis using 
 cosmic chronometers,  supernovae  type Ia, and  
Baryon Acoustic Oscillations data.
Concerning the Kaniadakis parameter, we find that it  is constrained 
around zero, namely around the value in which Kaniadakis entropy 
recovers standard Bekenstein-Hawking one. Additionally, for  the present matter 
density parameter
$\Omega_m^{(0)}$, we obtain a value slightly smaller compared to  $\Lambda$CDM scenario. 
Furthermore, we reconstruct the evolution of    the Hubble, 
deceleration and jerk parameters  extracting 
 the deceleration-acceleration transition redshift as 
 $z_T = 0.86^{+0.21}_{-0.14}$. Finally,  performing a detailed local and 
global dynamical system analysis,  we find that  the 
past attractor of the Universe is the matter-dominated solution, while the 
late-time stable solution is the  dark-energy-dominated one.
\end{abstract}


%%%%%%%%%%%%%%%%%%%%%%%%%%%%%%%%%%%%%%%%%%%%%%%%%%

%%%%%%%%%%%%%%%%% BODY OF PAPER %%%%%%%%%%%%%%%%%%

% The MNRAS class isn't designed to include a table of contents, but for this document one is useful.
% I therefore have to do some kludging to make it work without masses of blank space.
%\begingroup
%\let\clearpage\relax
%\tableofcontents
%\endgroup
%\newpage

%%%%%%%%%%%%%%%%%%%
\section{Introduction}
%%%%%%%%%%%%%%%%%%%

The acceleration of the Universe is one of the most  elusive problems in modern 
cosmology. Since its discovery in the last decade of the twentieth century by 
Supernovae (SNIa) observations \citep{Riess:1998,Perlmutter:1999}, and its confirmation 
by the acoustic peaks of the cosmic microwave background (CMB) radiation 
\citep{WMAP:2003elm}, it has been a theoretical and observational challenge to 
construct a model that combines all of its characteristics. From a theoretical 
point of view, and assuming homogeneous and isotropic symmetries (cosmological 
principle), the need for a component 
with features able to reproduce the Universe acceleration is vital to obtain 
accurate values for the observable Universe age and size. Recently, the 
confidence in the detection of this acceleration at late times has been 
increased with precise observations of the large scale structure 
\citep{Nadathur:2020kvq}. 

The best candidate to explain the observed acceleration 
is the well-known Cosmological Constant  (CC), interpreted  under the assumption that quantum 
vacuum fluctuations generate the constant energy density observed and, with this, 
a late-time acceleration. However, when we apply the Quantum Field Theory to 
assess the energy density, the result is in total discrepancy with 
observations, giving rise to the so-called {\it fine-tuning problem} 
\citep{Zeldovich:1968ehl, Weinberg}. In addition, recent observations developed by the collaboration \emph{Supernova $H_0$ for the Equation of State} (SH0ES) \citep{Riess:2020fzl} show  a 
discrepancy for the obtained value of $H_0$ when compared to Planck 
observations based on the $\Lambda$ Cold Dark Matter ($\Lambda$CDM) model
\citep{Aghanim:2018}. This generates a tension of $4.2\sigma$ between the 
mentioned experiments,   bringing a new crisis and the need for new ways to tackle 
the problem \citep{DiValentino:2020zio},  as long as this discrepancy is not related to unknown 
systematic errors  affecting the measurements 
\citep{DES:2019fny,birrer2021, Efstathiou:2021ocp, Freedman:2021ahq, 
Shah:2021onj}. Is in this vein that the community has been 
proposing other alternatives to  address  the problem of the Universe 
acceleration. In general, there are two main directions that one could follow. 
The first is to maintain general relativity an introduce new peculiar forms of 
matter, such as scalar fields \citep{Copeland:2006wr,Cai:2009zp,review:universe}, 
 Chaplygin gas \citep{Chaplygin,Villanueva_2015,Hernandez-Almada:2018osh}, 
viscous fluids \citep{Cruz, MCruz:2017, CruzyHernandez,AlmadaViscoso, 
Hernandez-Almada:2020ulm, Almada:2020}, etc, collectively known as dark-energy 
sector. The second way is to construct modified gravitational theories 
\citep{CANTATA:2021ktz,Capozziello:2011et} such as  braneworlds models
\citep{Maartens:2010ar,Garcia-Aspeitia:2016kak,Garcia-Aspeitia:2018fvw}, 
emergent gravity 
\citep{PEDE:2019ApJ,Pan:2019hac,PEDE:2020,Hernandez-Almada:2020uyr,
Garcia-Aspeitia:2019yni,Garcia-Aspeitia:2019yod}, Einstein-Gauss-Bonet 
\citep{Glavan:2019inb,Garcia-Aspeitia:2020uwq}, thermodynamical models 
\citep{Saridakis:2020cqq,Leon:2021wyx}, torsional gravity \citep{Cai:2015emx}, $f(R)$ theories \citep{Dainotti:2021},
etc.  

On the other hand, there is an  increasing interest in dark energy 
alternative models with the holographic principle. This is inspired by the 
relation  between entropy and  the area of a black hole.  It  states that the observable 
degree of   freedom  of a physical system in a volume can be encoded in a 
lower-dimensional description on its boundary \citep{Hooft:1993, Susskind:1995}. 
The holographic principle imposes a connection between the infrared (IR) 
cutoff, related to large-scale of the Universe, with the  ultraviolet (UV) one, 
related to the vacuum energy. Application of the holographic principle to the 
Universe horizon gives rise to   a vacuum energy   of holographic origin, namely holographic dark energy \citep{Li:2004rb,Wang:2016och}.  Holographic dark 
energy proves to lead to interesting phenomenology and, thus, it has been 
studied in detailed 
\citep{Li:2004rb,Wang:2016och,Horvat:2004vn, Pavon:2005yx,
Wang:2005jx,
Nojiri:2005pu,Kim:2005at,  Setare:2008pc,Setare:2008hm},  confronted to observations 
\citep{Zhang:2005hs,Li:2009bn,Feng:2007wn,Zhang:2009un,Lu:2009iv,
Micheletti:2009jy} and extended to various frameworks 
\citep{Gong:2004fq,Saridakis:2007cy,  
 Cai:2007us,Setare:2008bb,Saridakis:2007ns, 
Suwa:2009gm, BouhmadiLopez:2011qvd, 
Khurshudyan:2014axa,
 Saridakis:2017rdo,Nojiri:2017opc, 
Saridakis:2018unr,  
Kritpetch:2020vea,Saridakis:2020zol,Dabrowski:2020atl,
daSilva:2020bdc,   Mamon:2020spa,
Bhattacharjee:2020ixg,
Huang:2021zgj,Lin:2021bxv,Colgain:2021beg, 
Nojiri:2021iko,Shekh:2021ule}.

Recently,  an extension of the holographic dark energy scenario was constructed 
in \citep{Drepanou:2021jiv}, based on Kaniadakis entropy. The latter is an 
extended entropy arising from the relativistic extension of standard 
statistical theory, quantified by one new parameter  
\citep{Kaniadakis:2002zz,Kaniadakis:2005zk}. In the case where this Kaniadakis 
parameter becomes zero, i.e. when Kaniadakis entropy becomes the standard  
Bekenstein-Hawking 
entropy,  Kaniadakis-holographic dark energy recovers 
standard-holographic dark energy, however, in the general case, it exhibits  a range of   behaviors  with interesting cosmological implications.

In this work, we  investigate Kaniadakis-holographic dark energy, in 
order to tackle the late time universe acceleration  problem. The outline of the paper is 
as follows. In Section \ref{MB} the mathematical background of the model is 
considered, presenting the master equations.  Section \ref{sec:data} presents 
the observational confrontation analysis that includes three data samples and 
the results from the corresponding constraints. Section \ref{sec:SA} is 
dedicated to the dynamical system  investigation and the stability analysis. 
Finally, in Section \ref{sec:Con} we give a brief summary and a discussion of the 
results. Throughout the manuscript  we use natural units where
$\tilde{c}=\hbar=k_{B}=1$ (unless stated otherwise). 

%%%%%%%%%%%%%%%%%%%%%%%%%%%%%%%%%%%%%%%%%%%%%
\section{ Kaniadakis holographic dark energy} \label{MB}
%%%%%%%%%%%%%%%%%%%%%%%%%%%%%%%%%%%%%%%%%%%%%

In this section we briefly review Kaniadakis holographic dark energy and we 
elaborate the corresponding equations in order to bring them to a form suitable 
for observational confrontation. 
The essence of  holographic dark energy is the 
 inequality $\rho_{DE}  L^4\leq S$, with $\rho_{DE}$  being the holographic dark energy density, $L$  the largest distance 
(typically a horizon), and $S$ the entropy expression in the case of a black hole   with a horizon $L$  \citep{Li:2004rb,Wang:2016och}. In the   standard   
application using Bekenstein-Hawking entropy
$S_{BH}\propto A/(4G)=\pi L^2/G$, where $A$ is the area and $G$ the Newton's 
constant,  one obtains standard-holographic dark energy, 
i.e.  $\rho_{DE}=3c^2 M_p^2 L^{-2}$, where   $M_p^2=(8\pi G)^{-1}$ is the Planck mass 
and $c$ is the model parameter arising from the saturation of the above 
inequality.  
  
On the other hand, one can construct   the one-parameter
generalization of the classical   entropy, namely  Kaniadakis entropy     
$S_{K}=- k_{_B} \sum_i n_i\, \ln_{_{\{{\scriptstyle
K}\}}}\!n_i $ \citep{Kaniadakis:2002zz,Kaniadakis:2005zk},
where $k_{_B}$ is the Boltzmann constant   and
with $\ln_{_{\{{\scriptstyle
K}\}}}\!x=(x^{K}-x^{-K})/2K$. This is characterized by the   
dimensionless parameter $-1<K<1$, which  accounts for the relativistic
deviations from  standard statistical mechanics, and 
  in the limit $K\rightarrow0$ it recovers standard entropy.
  Kaniadakis entropy    can be re-expressed as 
\citep{Abreu:2016avj,Abreu:2017hiy,Abreu:2021avp}
\begin{equation}
 \label{kstat}
S_{K} =-k_{_B}\sum^{W}_{i=1}\frac{P^{1+K}_{i}-P^{1-K}_{i}}{2K},
\end{equation}
where  $P_i$  is the probability of  a specific microstate of the system and 
 $W$  the total number of possible configurations. Applied in the
   black-hole framework, it results   into 
\citep{Drepanou:2021jiv,Moradpour:2020dfm,Lymperis:2021qty}
 \begin{equation} \label{kentropy}
S_{K} = \frac{1}{K}\sinh{(K S_{BH})},
\end{equation}
which gives standard 
Bekenstein-Hawking entropy 
 in the limit  $K\rightarrow 0$.
 Finally, since any deviations from standard thermodynamics are expected to be 
small, one can approximate (\ref{kentropy}) for $K\ll1$, 
acquiring \citep{Drepanou:2021jiv}
\begin{equation}\label{kentropy2}
S_{K} = S_{BH}+ \frac{K^2}{6} S_{BH}^3+ {\cal{O}}(K^4).
\end{equation}
 
In order to analyze the dynamics of the universe, we consider the homogeneous and isotropic cosmology based on the 
Friedmann-Lema\^{i}tre-Robertson-Walker (FLRW) line element 
$ds^2=-dt^2+a(t)(dr^2+r^2d\Omega^2)$, where $d\Omega^2\equiv 
d\theta^2+\sin^2\theta d\varphi^2$, $a(t)$ is the scale factor and we consider 
null spatial curvature $k=0$.  Furthermore, as usual we use $L$ 
as the future event horizon $R_h\equiv a \int_t^{\infty }
   \frac{1}{a(s)} \, ds$. 
Inserting these into the above formulation, and using Kaniadakis entropy 
instead of Bekenstein-Hawking one, we extract the energy density of 
Kaniadakis holographic dark energy as \citep{Drepanou:2021jiv}
\begin{eqnarray}
&& \rho_{DE}= \frac{3c^2M_p^2}{R_h^2}+K^2 M_{p}^6
   R_h^2, \label{rhoDE}
\end{eqnarray}
with $c>0$ and $K$ being the two parameters of the  model. 
 Hence, we can write the   Friedmann and Raychaudhuri equations as
\begin{eqnarray}
    &&H^2=\frac{1}{3M_p^2}(\rho_m+\rho_{DE}), \label{Frie}\\
    &&\dot{H}=-\frac{1}{2M_p^2}(\rho_m+p_m+\rho_{DE}+p_{DE}), \label{Ray}
\end{eqnarray}
where $H\equiv \dot{a}/a$ is the Hubble parameter, $\rho_m$ and $p_m$ are the energy density and 
pressure of matter perfect fluid, while the matter conservation leads to dark 
energy conservation and, in turn, to the dark energy pressure
\begin{eqnarray}
   && p_{DE}=  -\frac{2c^2 M_p^2}{R_h^3 H}-\frac{
   c^2M_p^2}{R_h^2}+K^2 M_{p}^6\left[\frac{2R_h}{3
   H}-\frac{5}{3} R_h^2\right]. \label{pDE}
\end{eqnarray}
 
 The combination of Raychaudhuri equation \eqref{Ray} 
and  \eqref{rhoDE}, \eqref{pDE}  gives
\begin{eqnarray}
 \dot{H}&=&\frac{c^2}{R_h^3 H}+\frac{c^2(3
   w_m+1)}{2R_h^2}-\frac{3}{2}(w_m+1) H^2 \nonumber \\
   &&-K^2 M_p^4\left[\frac{R_h}{3
   H}-\frac{1}{6}
   R_h^2 (3 w_m+5)\right],
\end{eqnarray}
where $w_m\equiv p_m/\rho_m$ is the equation of state (EoS) parameter for 
matter, considered from now on as dust ($w_m=0$). From this expression  we can 
construct the deceleration and jerk parameters, which give us information about 
the transition to an accelerated Universe. Thus, using the definition 
of $R_h$,  
we obtain that the energy density is 
\begin{align}
    \rho_{DE} = & \frac{3 c^2 M_p^2}{a^2 \left(\int_t^{\infty } \frac{1}{a(s)} \,
   ds\right)^2}+K^2 M_{p}^6 a^2 \left(\int_t^{\infty }
   \frac{1}{a(s)} \, ds\right)^2, \label{2rhoDE}
\end{align}
and the pressure
\begin{align}
 p_{DE}= &  -\frac{2 c^2 M_p^2}{a^3 H
   \left(\int_t^{\infty } \frac{1}{a(s)} \, ds\right)^3}-\frac{c^2
   M_p^2}{a^2 \left(\int_t^{\infty } \frac{1}{a(s)} \,
   ds\right)^2}\nonumber \\
   & +K^2 M_{p}^6\left[\frac{2a \left(\int_t^{\infty }
   \frac{1}{a(s)} \, ds\right)}{3 H}  -\frac{5}{3}
   a^2 \left(\int_t^{\infty } \frac{1}{a(s)} \,
   ds\right){}^2\right]. \label{2pDE}
\end{align}
Moreover, the fractional energy density of DE is defined as
\begin{equation}
    \Omega_{DE}:=\frac{ \rho_{DE}}{3 M_p^2 H^2}= \frac{K^2 M_p^6 a^4 \left(\int_t^{\infty } \frac{1}{a(s)} \,
   ds\right)^4+3 c^2 M_p^2}{3 M_p^2 a^2 H^2
   \left(\int_t^{\infty } \frac{1}{a(s)} \, ds\right)^2}. \label{defODE1}
\end{equation}
From definition
\eqref{defODE1} we have four branches for 
\begin{equation}
\mathcal{I}(t):=     \int_t^{\infty }  {a(s)}^{-1} \, ds,
\end{equation}
which give four possible expressions for the particle  horizon
\begin{align}
 {R_h}_{1,2}(t)&= \mp  \frac{\left[3H^2\Omega_{DE}-\sqrt{9 H^4\Omega_{DE}^2-12 c^2 K^2
   M_p^4}\right]^{ {1}/{2}}}{\sqrt{2} |K| M_p^2}, \label{1y2}\\
{R_h}_{3,4}(t)&= \mp \frac{\left[3 H^2\Omega_{DE}+\sqrt{9 H^4\Omega_{DE}^2-12 c^2 K^2
   M_p^4}\right]^{ {1}/{2}}}{\sqrt{2} |K| M_p^2}. \label{3y4}
\end{align}
${R_h}_1(t)$ and ${R_h}_3(t)$ are both discarded since they lead to negative particle horizon. 
To decide between the choices ${R_h}_2(t)$ and ${R_h}_4(t)$, which  are both non negative, we calculate the limit $K\rightarrow 0$ and obtain 
\begin{equation}
\lim_{K\rightarrow 0} {R_{h}}_2 = \frac{c}{H
   \sqrt{\Omega_{DE}}}, \quad \lim_{K\rightarrow 0} {R_{h}}_4 = \infty.
\end{equation}
That is, $ {R_{h}}_2(t)$ defined by \eqref{1y2} is the only physical solution. 

%%VM quede aqui
We proceed by introducing   the usual dimensionless variable
\begin{equation}
    E\equiv \frac{H}{H_0},
    \label{Edefin}
\end{equation}
where $H_0$ is the Hubble constant at present time and, for convenience, we define the dimensionless constant $\beta\equiv \frac{K M_p^2}{H_0^2}$. 

Differentiating 
(\ref{defODE1}) and (\ref{Edefin}), and using the Friedmann equations we   obtain  
the master equations
\begin{align}
{\Omega_{DE}^{\prime}} =  (1-\Omega_{DE}) \left[3 (w_m+1) \Omega_{DE} +2  \mathcal{X} \right], \label{eq2.15}
\\
E^{\prime} =E \left[ -\frac{3}{2} (w_m+1) (1-\Omega_{DE})+  \mathcal{X}\right], \label{eq2.16}
\end{align}
     where
\begin{eqnarray}
 \mathcal{X} &\equiv &   \frac{1}{E^3}\left[\frac{2 \beta ^2 E^4 \Omega_{DE}^2-8 \beta ^4 c^2/3}{3 E^2 \Omega_{DE}-\sqrt{9 E^4 \Omega_{DE}^2-12 \beta ^2
   c^2}}\right]^{1/2}\nonumber \\
   &&-\frac{1}{E^2}[E^4 \Omega_{DE}^2-4 \beta ^2 c^2/3]^{1/2}. \label{eq2.17}
\end{eqnarray}
We use initial conditions $\Omega_{DE}(0)\equiv\Omega_{DE}^{(0)}=1- 
\Omega_m^{(0)}, E(0)=1$, 
where primes denote derivatives with respect to e-foldings number 
$N=\ln(a/a_0)$, and   $N=0$ marks the current time (from now on, the index 
``0'' marks the value of a quantity at present). 
The physical region of the phase space is 
\begin{equation}
  3  E^4 \Omega_{DE}^2 -4 \beta ^2 c^2\geq 0.
\end{equation}
Notice that $\mathcal{X}\rightarrow \frac{\Omega_{DE}^{ {3}/{2}}}{c}- \Omega_{DE}$
as $\beta\rightarrow 0$. 

From the matter conservation equation, we arrive at
\begin{equation}
\rho_m^{\prime}(N)= -3 (1+w_m)\rho_m, \quad \rho_m(0)= 3 M_p^2 H_0^2 \Omega_{m}^{(0)},
\end{equation}
and, therefore, we have $\rho_{m}(N)= 3 H_0^2M_p^2  \Omega_m^{(0)} e^{-3 N  (w_m+1)} $ which then  leads to
\begin{align}
   &  \Omega_{DE} (N)=1-   \Omega_{m} (N)= 1- \frac{\Omega_m^{(0)} 
e^{-3 N  (w_m+1)}}{E^2}.\label{defODE}
\end{align}
Defining $Z=E^2$, we obtain the equation 
\begin{equation}
Z^{\prime} = - 3 (w_m+1)  \Omega_m^{(0)} e^{-3 N  (w_m+1)} + 2 \mathcal{X} Z, \quad Z(0)=1, \label{final1}
\end{equation}
where 
\begingroup\makeatletter\def\f@size{7.5}\check@mathfonts
\begin{align}
  &\mathcal{X} Z=     -\left[ \left({{\Omega_m^{(0)}} e^{-3 N  (w_m+1)}}-{Z}\right)^2-\frac{4 \beta ^2 c^2}{3}\right]^{1/2} \nonumber \\
  & + \left[\frac{2 \beta ^2 \left(Z-{\Omega_m^{(0)}} e^{-3 N  (w_m+1)}\right)^2-\frac{8 \beta ^4 c^2}{3}}{3 Z^2-3 Z
   {\Omega_m^{(0)}} e^{-3 N  (w_m+1)} -Z\sqrt{9 \left(Z-{\Omega_m^{(0)}} e^{-3 N  (w_m+1)}\right)^2-12 \beta ^2 c^2}}\right]^{1/2}. \label{final2}
\end{align}
\endgroup
Thus, the evolution of $E^2(z)$ can be obtained by  substituting \eqref{final2} into \eqref{final1}.
More precisely, substituting \eqref{final2} into  \eqref{final1},  integrating, and imposing  
 the initial condition $Z(0)=1$, gives   $E^2(N)$. In order to express it as
$E^2(z)$, we use the relation $N= \ln (a/a_0)=-\ln(1+z)$, which is a relation 
between the e-folding ($N$), the scale factor ($a$), and the redshift ($z$).

Additionally, we can now write the deceleration parameter $q(z)$, and a cosmographic parameter which is 
related to the third-order derivative of the scale factor, i.e. the cosmographic jerk parameter $j(z)$, which are given by the formulas
\begin{align}
q :=& -1- \frac{E^{\prime}}{E}, \label{q}\\
j:= & q(2q+1)-q', \label{j}
\end{align}
where $j=1$ corresponds to the case of a  cosmological constant.

Hence, equation \eqref{q} becomes 
\begin{eqnarray}
    q&=&-1+\frac{3}{2} (w_m+1) (1-\Omega_{DE})- \mathcal{X},
\end{eqnarray}
with $ \mathcal{X}$ defined by \eqref{eq2.17}. 
$j$ is found by direct evaluation of \eqref{j}. 
We   have mentioned before  that  taking the limit $\beta \rightarrow 0$ in \eqref{eq2.15} 
and \eqref{eq2.16}, and neglecting error terms $O\left(\beta ^2\right)$, we acquire  the approximated differential 
equations
\begin{align}
    \Omega_{DE}^{\prime}= \frac{\Omega_{DE} (1-\Omega_{DE}) \left(3 w_m c+c+2 \sqrt{\Omega_{DE}}\right)}{c}, \label{ODEK0}
\end{align}
\begin{align}
E^{\prime}=  \frac{E \left\{2 \Omega_{DE}^{ {3}/{2}}+c [3 w_m 
(\Omega_{DE}-1)+\Omega_{DE}-3]\right\}}{2 c}. \label{HK0}
\end{align}
Equations \eqref{ODEK0} and \eqref{HK0} characterize standard holographic cosmology. 
Imposing the conditions
\begin{equation}
E(\Omega_{DE}^{(0)})=1, \;\;  \ln \left(\frac{a}{a_0}\right)\Big|_{\Omega_{DE}^{(0)}}=0,
\end{equation}
we obtain  the implicit solutions 
\begin{eqnarray}
E= &  \left(\frac{{\Omega_{DE}}}{{\Omega_{DE}^{(0)}}}\right)^{-\frac{3 (w_m+1)}{6 w_m+2}} \left(\frac{1-\sqrt{{\Omega_{DE}}}}{1-\sqrt{{\Omega_{DE}^{(0)}}}}\right)^{\frac{c-1}{3 c w_m+c+2}}
 \left(\frac{\sqrt{{\Omega_{DE}}}+1}{\sqrt{{\Omega_{DE}^{(0)}}}+1}\right)^{\frac{c+1}{3 c w_m+c-2}} \nonumber \\
   & \times \left(\frac{3 c w_m+c+2 \sqrt{{\Omega_{DE}}}}{3 c w_m+c+2 \sqrt{{\Omega_{DE}^{(0)}}}}\right)^{-\frac{12 (w_m+1)}{(3 w_m+1) \left((3 c w_m+c)^2-4\right)}},
\end{eqnarray}
and 
\begin{eqnarray}
&(1+z)^{-1}:= \left(\frac{a}{a_0}\right) \nonumber \\
&=  \left(\frac{\Omega_{DE}}{{\Omega_{DE}^{(0)}}}\right)^{\frac{1}{3 w_m+1}} \left(\frac{1-\sqrt{\Omega_{DE}}}{1-\sqrt{{\Omega_{DE}^{(0)}}}}\right)^{-\frac{c}{3 c w_m+c+2}} \left(\frac{\sqrt{\Omega_{DE}}+1}{\sqrt{{\Omega_{DE}^{(0)}}}+1}\right)^{-\frac{c}{3 c w_m+c-2}} \nonumber \\
 &  \times \left(\frac{3 c w_m+c+2 \sqrt{\Omega_{DE}}}{3 c w_m+c+2 \sqrt{{\Omega_{DE}^{(0)}}}}\right)^{\frac{8}{(3 w_m+1) \left((3 c
   w_m+c)^2-4\right)}}.
\end{eqnarray}
Lastly, expanding around $\beta=0$ and $\Omega_{DE}=1$ and removing second order terms,  the deceleration parameter \eqref{q} and the cosmographic jerk parameter \eqref{j} (in the dark-energy dominated epoch) are given by 
\begin{align}
&q =-\frac{1}{c}+\frac{(1-\Omega_{DE}) (3 c w_m+c+3)}{2 c}, \\
&j= \frac{2-c}{c^2}+\frac{(1- \Omega_{DE}) (3 c w_m+c+3) [c (3 w_m+2)-2]}{2 c^2}.
\end{align}
Furthermore, expanding around $\beta=0$ and $\Omega_{DE}=0$ and removing second order terms,  the deceleration parameter \eqref{q} and the cosmographic jerk parameter \eqref{j}  (in the matter dominated epoch) are given by 
\begin{align}
&q =\frac{1}{2} (3 w_m+1) (1- \Omega_{DE}),\\
&j=\frac{1}{2} [9 w_m (w_m+1)+2](1- \Omega _{DE}).
\end{align}

%%%%%%%%%%%%%%%%%%%%%%%%%%%%%%%%%%%%%%%%%%%%%%%%%%%%%%%%%%%%%%%%%%%%%%%%%%%%%%%%%%%%
%%%%%%%%%%%%%%%%%%%%%%%%%%%%%%%%%%%%%%%%%

\section{Observational analysis} \label{sec:data}

One of the goals of this work is to provide observational bounds on the 
parameter of Kaniadakis entropy $K$ or, more conveniently $\beta$, however we 
are also interested in the behavior of all cosmological parameters, namely on 
the vector
${\bm\Theta} = \{h, \Omega_m^{(0)}, \beta, c\}$. For 
the parameter estimation  we use the recent measurements of the observational 
Hubble data as well as data from type Ia supernovae, and baryon acoustic 
oscillations observations. In what follows, we first briefly introduce these 
datasets and the Bayesian methodology, and then we apply it in the scenario of 
Kaniadakis-holographic dark energy,  providing the resulting observational 
constraints.  

\subsection{Data and methodology } \label{sec:Data}
 
%%%%%%%%%%%%%%%%%%%%%%%%%%%%%%%%%%%%%%%%%
\subsubsection{Cosmic chronometer data}

The Hubble parameter $H(z)$ describes the expansion rate of the Universe as a 
function of redshift $z$. Currently, this parameter can be estimated from baryon 
acoustic oscillations measurements and differential age in passive galaxies 
(dubbed as cosmic chronometers).  While   the former could be biased due to 
the assumption of a fiducial cosmology, the samples from cosmic  
chronometers are independent from the underlying cosmological model. Thus, in 
this work we only consider the $31$ points from cosmic chronometer sample 
presented in \citet{Moresco:2016mzx,Magana:2018} in the redshift range 
$0.07<z<1.965$. We assume a Gaussian likelihood function for this observation 
as $\mathcal{L}_{\mathrm {CC}}\propto \exp{(-\chi_{\mathrm{CC}}^{2}/2)}$, where 
the figure-of-merit is
 \begin{equation} \label{eq:chi2_CC}
 \chi^2_{\mathrm{CC}}=\sum_i^{31}\left[\frac{H_{mod}({\bm\Theta},z_i)-H_{dat}(z_i)}{\sigma^i_{dat} }\right]^2,
\end{equation}
where $H_{dat}(z_i)$  and $\sigma^i_{obs}$ are the measured Hubble parameter and its observational uncertainty 
at the redshift $z_i$, 
respectively. The 
predicted Hubble parameter by the Kaniadakis-holographic dark energy is denoted 
by $H_{mod}({\bm\Theta})$, and it  can be obtained by solving the system 
of equations   \eqref{eq2.15}-\eqref{eq2.17}.

%%%%%%%%%%%%%%%%%%%%%%%%%%%%%%%%%%%%%%%%%
\subsubsection{Pantheon SNIa sample} 
    
Since the discovery of the late cosmic  acceleration with the observations of 
high redshift type Ia supernovae  (SNIa) by \citet{Riess:1998, Perlmutter:1999}, 
the 
observation of these distant objects is a crucial test to determine if a 
cosmological scenario is a viable candidate for the description of the 
late-time Universe. The probe consists of confronting the observed luminosity 
distance (or distance module) of SNIa with the theoretical prediction of any 
model. Up %today,
to now, the Pantheon sample \citep{Scolnic:2018} is the largest 
collection of high-redshift SNIa, with $1048$ data points with measured redshifts 
%measured
in the range $0.001<z<2.3$. The authors also provide a binned sample 
containing 40 points of binned distances $\mu_{dat, bin}$ in the redshift range 
$0.014<z<1.61$. In this work, we use the binned set and we consider a 
Gaussian likelihood $\mathcal{L}_{SNIa} \propto \exp{(-\chi_{SNIa}^2/2)}$. By 
marginalizing the nuisance parameters, the figure-of-merit function 
$\chi_{SNIa}^2$ is given by  
\begin{equation}
\chi_{SNIa}^{2}=a +\log \left( \frac{e}{2\pi} \right)-\frac{b^{2}}{e}, \label{fPan}
\end{equation}
where 
$a=\Delta\boldsymbol{\tilde{\mu}}^{T}\cdot{\bm C_{P}^{-1}}
\cdot\Delta\boldsymbol{\tilde{\mu}},\, 
b=\Delta\boldsymbol{\tilde{\mu}}^{T}\cdot\bm{C_{P}^{-1}}\cdot\Delta{\bm 1}$, $e=\Delta{\bm 1}^{T}\cdot{\bm C_{P}^{-1}}\cdot\Delta{\bm 1}$, and 
$\Delta\boldsymbol{\tilde{\mu}}$ is the vector of residuals between the model 
distance modulus and the observed  (binned) one. The covariance matrix $\bm{C_{P}}$ 
takes into account systematic and statistical uncertainties \citep{Scolnic:2018}.
Moreover, the theoretical counterpart of the distance modulus for any 
cosmological model is given by $\mu_{mod}({\bm\Theta},z) = 5 \log_{10} \left( 
d_L({\bm\Theta},z) / 10 {\rm pc} \right)$, where $d_{L}$ is the luminosity 
distance given by
 \begin{equation}
d_{L}({\bm\Theta},z)=\frac{\tilde{c}}{H_{0}}(1+z) \int^{z}_{0}\frac{{\rm dz}^{\prime}}{E(z^{\prime})},
\label{eq:dl}
\end{equation}
where $\tilde{c}$ is the light speed.

%%%%%%%%%%%%%%%%%%%%%%%%%%%%%%%%%%%%%%%%%
\subsubsection{Baryon Acoustic Oscillations}

Baryon Acoustic Oscillations (BAO) are fluctuation patterns in the matter density field as result of  internal interactions  in the hot primordial plasma  during the pre-recombination stage. Based on luminous 
red galaxies, a sample of 15 transversal BAO scale measurements within the 
redshift $0.110<z<2.225$ were collected by \citet{Nunes_2020}. %By
Assuming a 
Gaussian likelihood, $\mathcal{L}_{\mathrm {BAO}}\propto 
\exp{(-\chi_{\mathrm{BAO}}^{2}/2)}$, we build the figure of merit function as
\begin{equation} \label{eq:chi2_BAO}
\chi^2_{\rm BAO} = \sum_{i=1}^{15} \left[ \frac{\theta_{dat}^i - 
\theta_{mod}(\Theta,z_i) }{\sigma_{\theta_{dat}^i}}\right]^2\,,
\end{equation}
where $\theta_{dat}^i \pm \sigma_{\theta_{dat}^i}$ is the  BAO angular scale and 
its uncertainty at $68\%$ measured at $z_i$. The theoretical BAO angular scale 
counterpart, denoted as $\theta_{mod}$, is estimated by
\begin{equation}
    \theta_{mod}(z) = \frac{r_{drag}}{(1+z)D_A(z)}\,,
\end{equation}
where $D_A=d_L(z)/(1+z)^2$ is the angular diameter distance at $z$ which  
depends on the dimensionless luminosity distance $d_L(z)$, and $r_{drag}$ is 
the sound horizon at the baryon drag epoch,  considered to be $r_{drag}=137.7 \pm 
3.6\,$Mpc \citep{Aylor:2019}.

\subsubsection{Bayesian analysis} 

A Bayesian statistical  analysis based on Markov Chain Monte Carlo (MCMC) 
algorithm is performed to bound the free parameters of the Kaniadakis-holographic dark energy. The MCMC approach is implemented through the \texttt{emcee} python 
module \citep{Emcee:2013} in which we generate $1000$ chains with $250$ steps,  
each one after a burn-in phase. The latter is stopped when the chains have 
converged based on the auto-correlation time criteria. Thus, the inference of 
the parameter space is obtained by minimizing a Gaussian log-likelihood, 
$-2\ln(\mathcal{L}_{\rm data})\varpropto \chi^2_{\rm data}$, considering flat 
priors in the intervals: $h\in[0.2,1]$,  $\Omega_m^{(0)}\in[0,1]$,  
$\beta\in[-1,1]$,  $c\in[0,2]$ for each dataset.
Additionally, a combined analysis is performed by assuming no correlation 
between the datasets, hence the figure of merit is 
\begin{equation} \label{eq:chi2_joint}
\chi^2_{\rm Joint} = \chi^2_{\rm CC}+\chi^2_{\rm SNIa}+\chi^2_{\rm BAO}\,,
\end{equation}
namely,  the sum of the $\chi^2$ corresponding to each sample as previously defined. 
%previously.

%%%%%%%%%%%%%%%%%%%%%%%%%%%%%%%%%%%%%%%%%
\subsection{Results from observational constraints}

\begin{figure}
    \centering
    \includegraphics[width=0.5\textwidth]{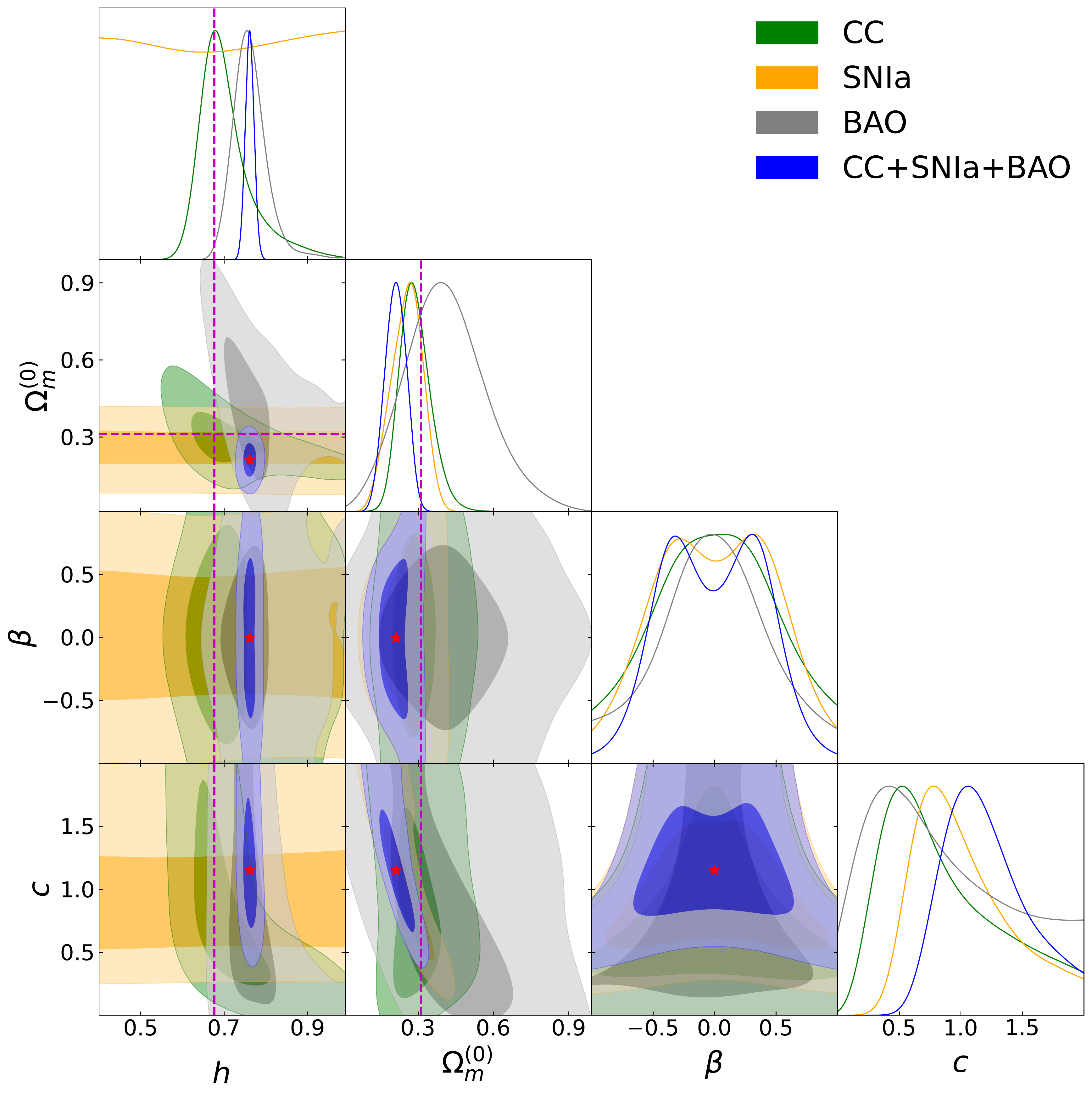}
    \caption{
    Two-dimensional likelihood contours at $68\%$ and $99.7\%$  confidence level  
(CL), alongside the corresponding 1D posterior distribution of the free 
parameters, in Kaniadakis-holographic dark energy case. The stars 
denote the mean values using the joint analysis,  and the dashed lines represent 
the best-fit values for $\Lambda$CDM cosmology \citep{Planck:2020}.  }
 \label{fig:contours}
\end{figure}

We perform the full confrontation described above for the scenario  of 
Kaniadakis holographic dark energy, and in 
Fig.  \ref{fig:contours} we present the 2D parameter likelihood contours at 
$68\%$ ($1\sigma$) and $99.7\%$ ($3\sigma$) confidence level (CL) respectively, 
alongside the corresponding 1D posterior distribution of the parameters. 
Additionally, Table \ref{tab:bestfits} shows the mean values of the parameters 
and their uncertainties at $1\sigma$.

%%%%%%%%%%%%%%%%%%%%%%%%%%%%%%%%%%%%%%%%%
\begin{table*}
\caption{Mean values of various parameters and their $68\%$ CL 
uncertainties for Kaniadakis-holographic dark energy. The quantities $\Delta$AICc ($\Delta$BIC) are the 
differences with respect to $\Lambda$CDM paradigm.}
\centering
%\resizebox{\textwidth}{!}{%
\begin{tabular}{|lccccccc|}
\hline
Sample     &    $\chi^2$     &  $h$ & $\Omega_m^{(0)}$ & $\beta$ & $c$ & 
$\Delta$AICc  & $\Delta$BIC \\
\hline 
CC & $14.69$ & $0.690^{+0.072}_{-0.043}$  & $0.284^{+0.066}_{-0.055}$  & $0.012^{+0.486}_{-0.489}$  & $0.729^{+0.665}_{-0.350}$  & $5.2$ & $7.0$ \\ [0.9ex] 
SNIa & $48.52$ & $0.597^{+0.279}_{-0.271}$  & $0.259^{+0.059}_{-0.069}$  & $0.013^{+0.480}_{-0.482}$  & $0.932^{+0.492}_{-0.302}$ & $5.1$ & $7.5$ \\ [0.9ex] 
BAO & $13.01$ & $0.758^{+0.041}_{-0.035}$  & $0.403^{+0.167}_{-0.151}$  & $-0.006^{+0.433}_{-0.418}$  & $0.756^{+0.759}_{-0.463}$ & $8.2$ & $5.6$ \\ [0.9ex] 
CC+SNIa+BAO & $98.07$ & $0.761^{+0.011}_{-0.010}$  & $0.211^{+0.043}_{-0.044}$  & $-0.003^{+0.412}_{-0.420}$  & $1.151^{+0.401}_{-0.287}$ & $21.7$ & $26.1$ \\ [0.9ex] 
\hline
\end{tabular}
%}
\label{tab:bestfits}
\end{table*}

\begin{figure}
    \centering
    \includegraphics[width=0.5\textwidth]{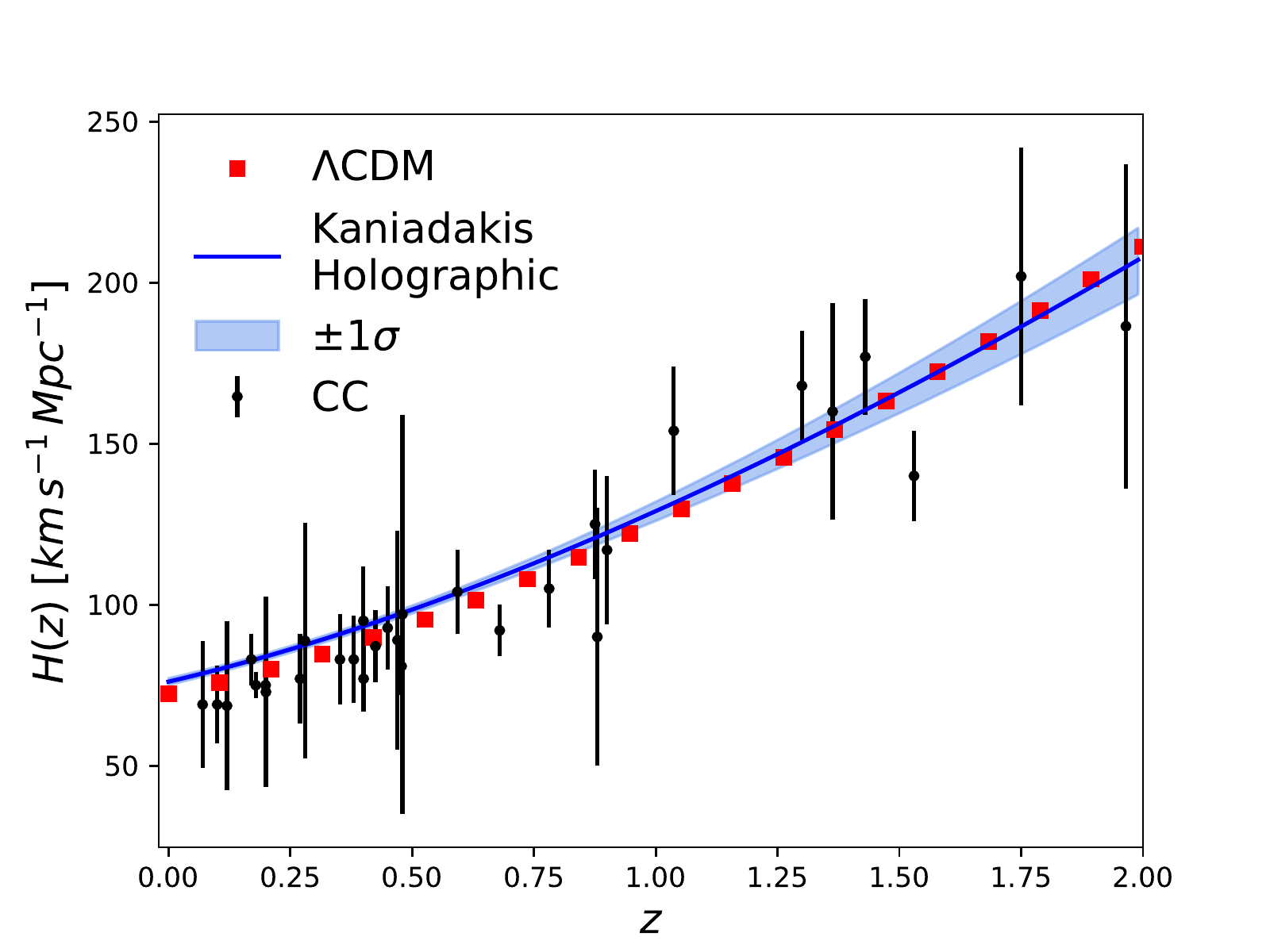}\\
    \includegraphics[width=0.5\textwidth]{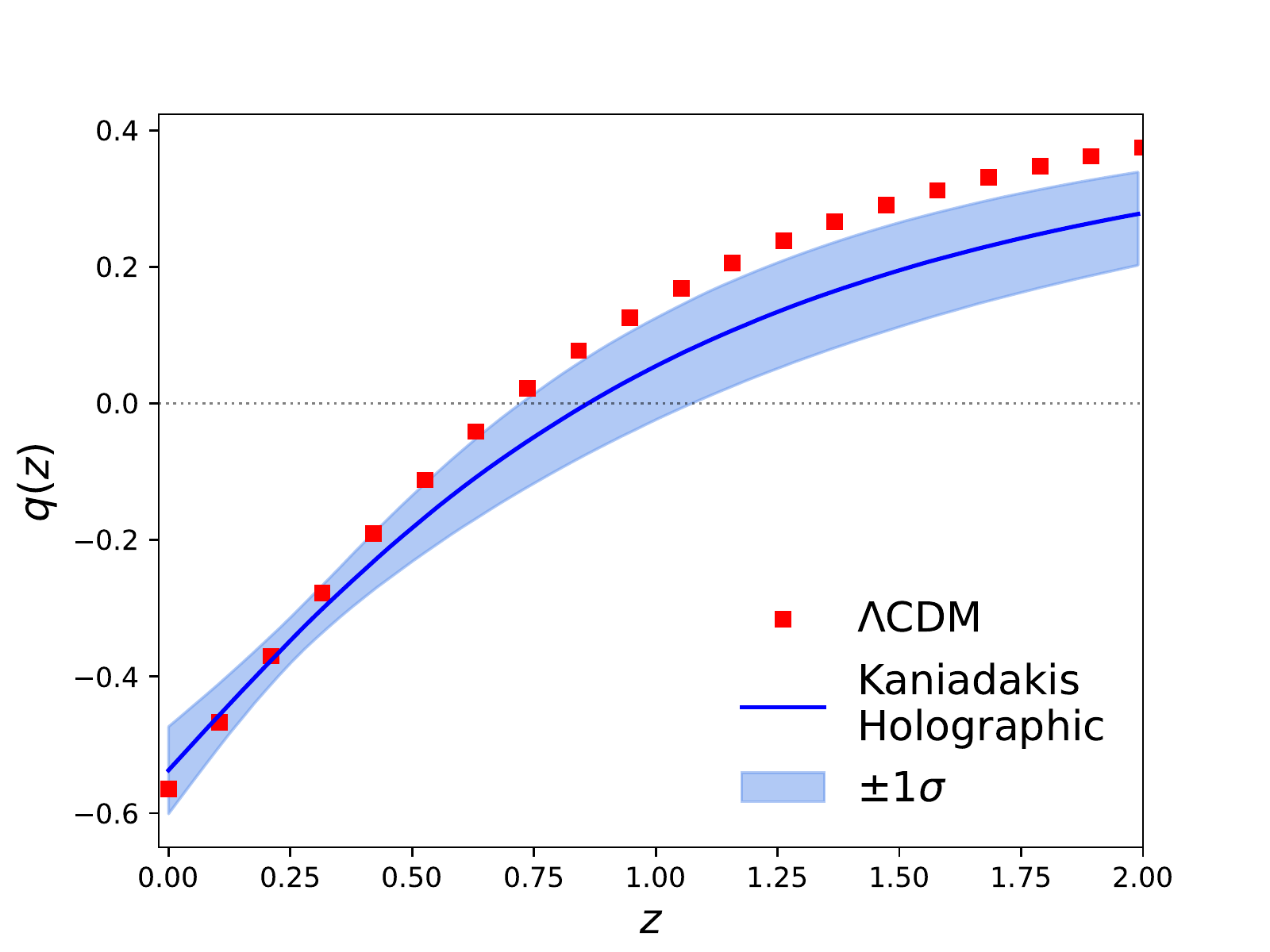}\\
    \includegraphics[width=0.5\textwidth]{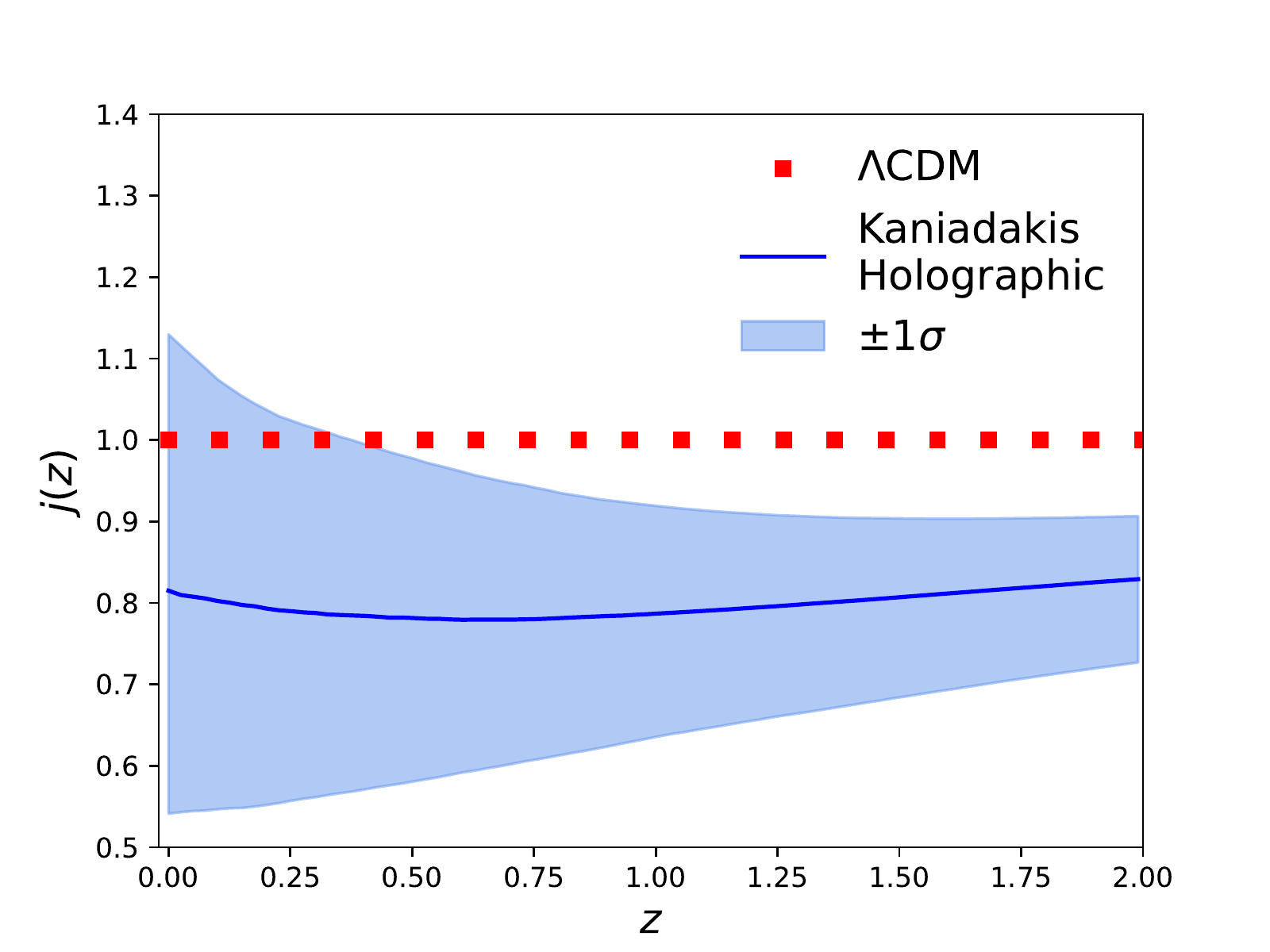}
    \caption{Reconstruction of %The
    the Hubble function  ($H(z)$, upper panel), the deceleration 
parameter ($q(z)$, middle panel), and the jerk parameter  ($j(z)$, bottom panel)
for the
Kanadiakis-holographic dark energy using the combined (CC+SNIa+BAO) analysis in the redshift range $0<z<2$. 
%In all panels 
The shaded regions 
represent the $68\%$ confidence level, 
and the square points
depict  the results of the 
    $\Lambda$CDM scenario with $h=0.723$ and $\Omega_m^{(0)}=0.290$, namely the values obtained through observational confrontation using the same datasets with the analysis of Kaniadakis holographic dark energy.}
    \label{fig:cosmography}
\end{figure}

In order to statistically compare these results 
with 
$\Lambda$CDM cosmology,  we apply the corrected 
Akaike information criterion (AICc) \citep{AIC:1974, Sugiura:1978, AICc:1989} and 
the Bayesian information criterion (BIC) \citep{schwarz1978}. 
They give a 
penalty according to size of data sample ($N$) and the  number of degrees of freedom 
($k$) defined as 
${\rm AICc}= \chi^2_{min}+2k +(2k^2+2k)/(N-k-1)$ 
and ${\rm BIC}=\chi^2_{min}+k\log(N)$ respectively, where $\chi^2_{min}$ is the 
minimum value of the $\chi^2$.  Thus, a model with lower values of AICc and BIC 
is preferred by the data. According to the difference between a given model and 
the reference one, denoted as $\Delta\rm{AICc}$, one has the following: 
if $\Delta \rm{AICc}<4$, both models are supported by the data equally, i.e 
they are statistically equivalent.
If $4<\Delta\rm{AICc}<10$, the data still support the given model but less than the preferred one.
If $\Delta \rm{AICc}>10$,
it indicates that the data does not support the 
given model.
Similarly, the difference between a candidate model and the reference model,  
denoted as $\Delta \rm{BIC}$, 
is interpreted 
in this way: 
if $\Delta \rm{BIC}<2$,
there is no evidence against the candidate 
model,
if $2<\Delta\rm{BIC}<6$, %indicates  that 
there is modest evidence against the candidate model,
if $6<\Delta \rm{BIC}<10$, there is strong evidence against the candidate model, and
$\Delta \rm{BIC}>10$ gives the strongest evidence against it. 
Hence, we have performed the above comparison, taking $\Lambda$CDM scenario as 
the reference model, and
we display the results in the last two columns of
Table \ref{tab:bestfits}.

A first observation is that the Kaniadakis parameter $\beta$ is constrained 
around 0 as expected, namely around the value in which Kaniadakis entropy 
recovers the standard Bekenstein-Hawking one. A second observation is that 
the scenario at hand gives a slightly smaller value for $\Omega_m^{(0)}$ 
comparing to  $\Lambda$CDM cosmology, however it
estimates 
a higher value for the 
present Hubble constant $h$, closer to its direct measurements through long-period Cepheids.
In particular, it is  consistent within  
$1\sigma$ with the value reported by \citet{Riess:2019} and it exhibits a 
deviation of  $4.18\sigma$ from the one obtained by Planck \citet{Planck:2020}.
On the other hand, based on our mean value of $c=1.151^{+0.401}_{-0.287}$ it is interesting that we do not observe a turning point in the $H(z)$ reconstruction shown in Fig. \ref{fig:cosmography}, a feature from which the usual holographic dark energy suffers when $c<1$ \citep{Colgain:2021beg}. Hence, we deduce that Kaniadakis holographic dark energy
can also solve such a problem and thus avoid to violate the Null Energy Condition (NEC).

Concerning the comparison with $\Lambda$CDM scenario, 
for the combined dataset analysis we find that $\Delta\rm{AICc}$ 
implies that $\Lambda$CDM is strongly favored 
over Kaniadakis-holographic dark energy. 
This result is also supported by BIC, for which $\Delta \rm{BIC}$  gives a strong 
evidence against it. Notice that these comparisons were performed by using the same datasets for both models $\Lambda$CDM and Kaniadakis cosmology.

Finally, based on the combined (CC+SNIa+BAO) analysis, in Fig. 
\ref{fig:cosmography} we present the reconstruction of the Hubble parameter 
$H(z)$, %of 
the deceleration parameter $q(z)$ (equation \eqref{q}), and 
the cosmographic jerk parameter
$j(z)$ (equation \eqref{j}), in the redshift range $0<z<2$. For comparison, we also depict  the 
corresponding curves for   $\Lambda$CDM scenario. 
Concerning the current values, our analysis leads to 
$H_0 = 76.09^{+1.06}_{-1.02}\, \rm{km/s/Mpc}$, 
$q_0 = -0.537^{+0.064}_{-0.064}$, 
$j_0 = 0.815^{+0.315}_{-0.274}$, where the uncertainties correspond to 
$1\sigma$ CL.  Additionally, using the joint analysis we find the redshift
for the deceleration-acceleration transition as $z_T = 
0.860^{+0.213}_{-0.138}$, and the Universe age as $t_U = 
13.000^{+0.406}_{-0.350} \,\rm{Gyrs}$. 
Notice that $z_T$ value is 
in agreement within $1\sigma$ with the value reported in 
\citet{Herrera-Zamorano:2020} for $\Lambda$CDM paradigm ($z_T=0.642^{+0.014}_{-0.014}$).

%%%%%%%%%%%%%%%%%%%%%%%%%%%%%%%%%%%%%%%%%%%%%
\section{Dynamical system and stability analysis} \label{sec:SA}
%%%%%%%%%%%%%%%%%%%%%%%%%%%%%%%%%%%%%%%%%%%%%

In this section we apply the powerful method of phase-space and stability 
analysis, which allows us to obtain a qualitative description of the local and 
global dynamics of cosmological scenarios, independently of the initial 
conditions and the
specific evolution of the universe. 
The extraction of asymptotic solutions give theoretical values that can be compared with the observed ones, such as the dark-energy and total equation-of-state parameters, the deceleration parameter, the density parameters of the different sectors, etc., and also allows the classification of the cosmological solutions \citep{Ellis}.

In order to perform the stability analysis of a given cosmological scenario,
one  transforms it to its autonomous form $\label{eomscol}
\textbf{X}'=\textbf{f(X)}$
\citep{Ellis,Ferreira:1997au,Copeland:1997et,Perko,Coley:2003mj,Copeland:2006wr,Chen:2008ft,Cotsakis:2013zha,Giambo:2009byn},
where $\textbf{X}$ is the column vector containing the auxiliary variables and 
primes denote derivative
with respect to a conveniently chosen time variable. Then, one extracts the  critical points
$\bm{X_c}$  by imposing the condition  $\bm{X}'=0$ and, %in order 
to determine
their stability properties,  one expands around them with $\textbf{U}$ the
column vector of the perturbations of the variables. Therefore,
for each critical point the perturbation equations are expanded to first
order as $\label{perturbation} \bm{U}'={\bm{Q}}\cdot
\bm{U}$, with the matrix ${\bm {Q}}$ containing the coefficients of the
perturbation equations. Finally, the eigenvalues of ${\bm {Q}}$ determine the 
type and stability of the   critical point under consideration.

%%%%%%%%%%%%%%%%%%%%%%%%%%%%%%%%%%%%%%%%%
\subsection{Local dynamical system formulation}

In this subsection we study the stability of system 
\eqref{eq2.15}-\eqref{eq2.16} with $\mathcal{X}$ defined in \eqref{eq2.17},  in 
the phase space 
\begin{equation}
 \left\{(E, \Omega_{DE})\in \mathbb{R}^2:  3  E^4 \Omega_{DE}^2 -4 \beta ^2 c^2\geq 0 \right\}.\label{Phase35}
\end{equation}
For generality, we keep the matter equation-of-state parameter $w_m$ in the 
calculations, and it can be set to zero in the 
final result if needed. 
Since $\beta$ and $c$ appear quadratic in 
\eqref{eq2.15}, \eqref{eq2.16} 
\eqref{eq2.17} and \eqref{Phase35}, these equations are invariant under the changes $c\mapsto -c$ and  $\beta\mapsto -\beta$. Therefore, in this section  we focus on $\beta>0$ and $c>0$. When $\beta<0$ we change $\beta$ by $-\beta$ and $c$ by $-c$ on the next discussion.

The equilibrium points dominated by dark energy (namely possessing 
$\Omega_{DE}=1$) with finite $H$  are:
\begin{itemize}
    \item $L_1: (E, \Omega_{DE})=\left(\frac{\sqrt{2  \beta  c }}{\sqrt[4]{3}},1\right)$. This point always satisfies 
    $-12 c^2\beta^2 + 
9 E^4\Omega_{DE}^2 =0$.  The eigenvalues are $\left\{-3 (w_m+1),\infty \; 
\text{sgn}\left(\left(\sqrt{2}-2 c\right)\right)\right\}$. It is a stable point 
for $c>\frac{\sqrt{2}}{2}$ and $w_m>-1$, and a saddle for 
$c<\frac{\sqrt{2}}{2}$ and $w_m>-1$ .  
    
    \item  $L_2: (E, \Omega_{DE})=\left(\frac{\sqrt{ \beta }}{\sqrt[4]{3(1-c^2)}}, 1\right)$. This point satisfies the reality condition if 
$\frac{3 \beta ^2 \left(1-2 c^2\right)^2}{1-c^2}\geq 0$, namely $\beta =0, c^2> 
1$ or $\beta\neq 0, c^2<1$. For $ c^2\leq \frac{1}{2}$ the eigenvalues are
\begin{eqnarray*}
{\lambda_1, \lambda_2}= \left\{\frac{ \left(4 c^4-4 c^2-1\right) |c|+\left(-8 
c^4+6 c^2+1\right)
   \sqrt{  1-c^2}}{  \left|c-2 c^3\right|},  \right.\nonumber \\
   \left.
   2
   \left(\sqrt{\frac{1}{c^2}-1}-1\right) \left(2 c^2-1\right)-3
   (w_m+1)\right\}.
\end{eqnarray*}
This is a saddle point,  as it can be 
verified numerically in Fig. \ref{fig:StabilityL2}.   
 Moreover, for $\frac{1}{2}<c^2<1$, the eigenvalues are $\left\{2-2 c^2,-3 
(w_m+1)\right\}$, and thus for   $w_m>-1$  it is also a 
saddle point. 
   \begin{figure}
       \centering
       \includegraphics[width=0.5\textwidth]{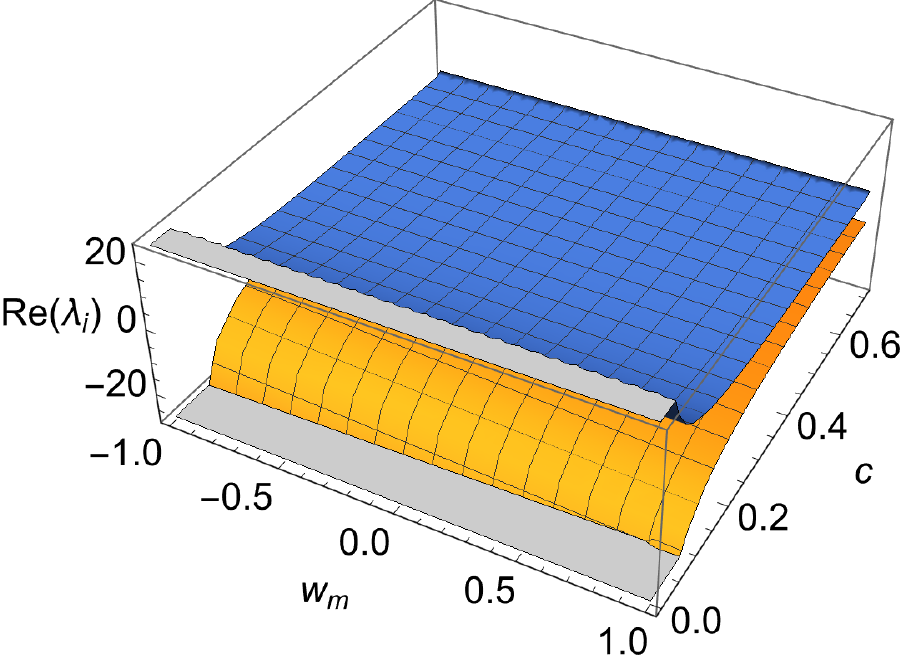}
       \caption{{\it{The eigenvalues corresponding to the point  $L_2$, for 
$w_m\in[-1,1]$, $c\in [0, \sqrt{2}/2]$. }}}
       \label{fig:StabilityL2}
   \end{figure}
  \end{itemize}
Since $\Omega_{DE}^2 \geq \frac{4 \beta ^2 c^2}{3E^4}\geq 0$, we deduce that 
the only possibility to have matter domination, namely $\Omega_{DE}=0$, is when 
$E\rightarrow \infty$, due to the reality condition $c^2 \beta^2 \geq 0$. 
It is convenient to define the dimensionless compact variable $T=(1+E)^{-1}$
such that $T\rightarrow 0$ as $E\rightarrow \infty$ and $T\rightarrow 1$ as $E\rightarrow 0$.  Then, we obtain 
\begin{small}
\begin{eqnarray}
&
\!\!\!\!\!\!\!\!\!\!\!\!\!\!\!\!\!\!\!\!\!\!
T'= \frac{3}{2} (T-1) T (w_m+1)
   (\Omega_{DE}-1) -\frac{T^3
   \sqrt{\frac{(T-1)^4 \Omega_{DE}^2}{T^4}-\frac{4 \beta ^2 c^2}{3}}}{T-1} \nonumber \\
   & \!\!\!\!\!\!\!\! -\frac{ T^5 \sqrt{\frac{2\beta^2 (T-1)^4 
\Omega_{DE}^2}{T^4}-\frac{8 \beta ^4 c^2}{3}}}{(T-1)^2 \sqrt{3
   (T-1)^2 \Omega_{DE}-\sqrt{9 (T-1)^4 \Omega_{DE}^2-12 \beta ^2 c^2 T^4}}}, \label{syst1}
   \end{eqnarray}
\begin{eqnarray}
&
\!\!\!\!\!\!\!\!\!\! \!\!\!\!\!\!\!\!\! \! \! \! \! \!  
\! \! \! \! 
\Omega_{DE}'= (\Omega_{DE}-1) \left[    -3 (w_m+1) \Omega_{DE} + \frac{2 T^2 
\sqrt{\frac{(T-1)^4 \Omega_{DE}^2}{T^4}-\frac{4 \beta ^2 c^2}{3}}}{(T-1)^2} 
\right. \nonumber \\
   &
   \ \ \ \ \ \ \ \ \ \ \ \ 
   \left. +\frac{2  T^2 \sqrt{2  \beta^2 (T-1)^4 \Omega_{DE}^2-\frac{8}{3} \beta 
^4 c^2 T^4}}{(T-1)^3 \sqrt{3 (T-1)^2 \Omega_{DE}-\sqrt{9 (T-1)^4 
\Omega_{DE}^2-12 \beta ^2 c^2 T^4}}}\right], \label{syst2}
\end{eqnarray}
\end{small}
defined on the physical region 
\begin{equation}
    9 (T-1)^4 \Omega_{DE}^2-12 \beta ^2 c^2 T^4\geq 0.
\end{equation}

In summary the sources/sinks are:
\begin{itemize}
    \item $L_1: (E, \Omega_{DE})=\left(\frac{\sqrt{2  \beta  c }}{\sqrt[4]{3}},1\right)$  is a stable point 
for $c>\frac{\sqrt{2}}{2}$ and $w_m>-1$, and a saddle for 
$c<\frac{\sqrt{2}}{2}$ and $w_m>-1$.  
    \item For the dark-energy dominated solution $L_3: (T, \Omega_{DE})=(0,1)$, the eigenvalues are 
$\left\{\frac{c-1}{c},-\frac{3 c w_m+c+2}{c}\right\}$, thus  it is a stable 
point for $-1<w_m<1$ and  $ 0<c<1$ or a saddle for  $-1<w_m<1$ and  $ c>1$.
    
    \item The past attractor is the matter dominated solution $L_4: (T, \Omega_{DE})=(0,0)$, for which the eigenvalues are         
$\left\{3 (w_m+1),\frac{3 (w_m+1)}{2}\right\}$, and since they are always 
positive for $-1<w_m<1$ it is an unstable point. 
\end{itemize}

We remark here that $E=E_c$ finite corresponds to the de Sitter solution with $H= 
E_c H_0$, and $a(t)\propto e^{ E_c H_0 t}$. That is, point $L_1$ satisfies  $a(t)\propto e^{ \frac{\sqrt{2 |\beta c|}}{\sqrt[4]{3}} H_0 t}$  and 
it is a late-time attractor providing the  accelerated regime. Additionally, for $\beta\neq 0, c^2<1$, the point $L_2$ exists and satisfies  $a(t)\propto e^{\frac{\sqrt{|\beta| }}{\sqrt[4]{3 (1-c^2)}} H_0 t}$, and since it is a saddle it can provide a transient   accelerated   phase  that can be related to inflation.

\begin{figure}
    \centering
    \includegraphics[width=8cm,scale=0.6]{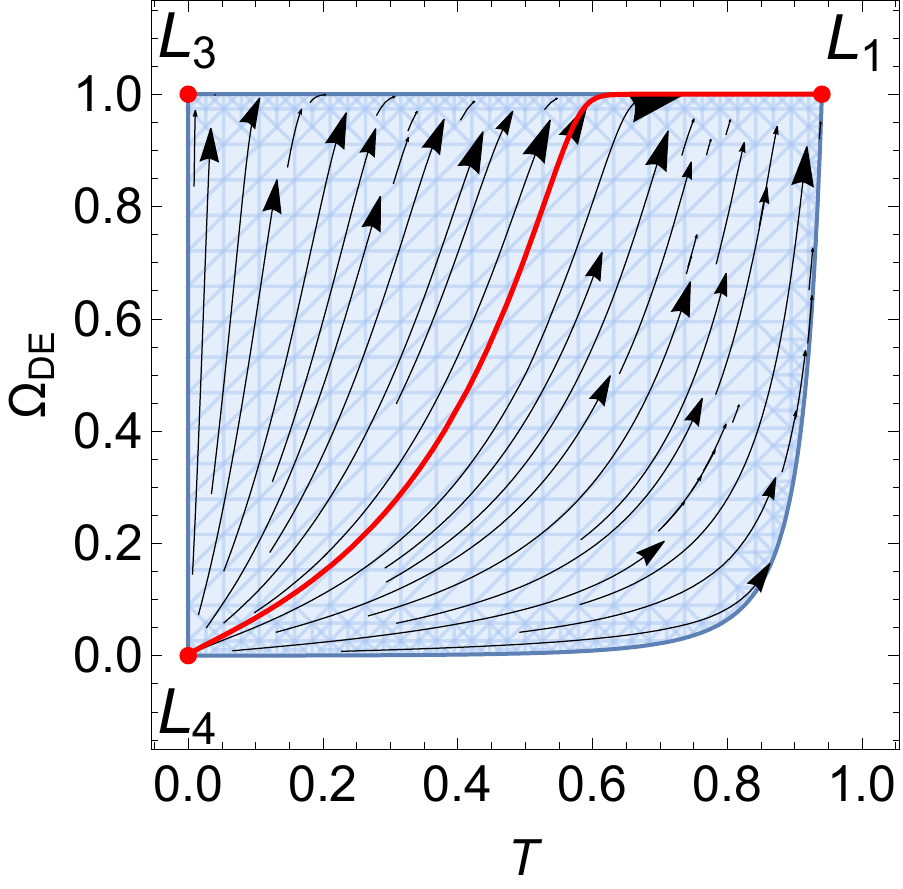}
    \caption{{\it{Phase-space plot of the dynamical system 
\eqref{syst1}-\eqref{syst2}, for the best fit values $\beta = -0.003$ and $c = 1.151$ of Kaniadakis-holographic dark energy, and for dust matter ($w_m = 0$). 
The red curve represents the solution for the initial data 
$\Omega_{DE}|_{z=0}=0.71$, corresponding to the mean value from the joint analysis CC+SNIa+BAO,  and for $T|_{z=0}=0.5$. The 
dashed blue region is the physical region  $9 (T-1)^4 \Omega_{DE}^2-12 \beta ^2 
c^2 T^4\geq 0$, where the equations are real-valued. }}}
    \label{DS1}
\end{figure}

In order to present the results in a more transparent way, in Fig. \ref{DS1} we show 
a phase-space plot of the system \eqref{syst1}-\eqref{syst2} for the 
best fit values $\beta = -0.003$ and $c = 1.151$ and for dust matter ($w_m = 
0$). The red curve represents the solution for the initial data 
$\Omega_{DE}|_{z=0}=0.71$, corresponding to the mean value from the joint analysis CC+SNIa+BAO,  and for $T|_{z=0}=0.5$. The dashed blue region is 
the physical region  $9 (T-1)^4 \Omega_{DE}^2-12 \beta ^2 c^2 T^4\geq 0$, where 
the equations are real-valued. 
From this figure it is confirmed that the late-time attractor is the 
dark-energy dominated solution $\Omega_{DE}=1$ with $T=0$. The past attractor 
is the matter-dominated solution $\Omega_{DE}=0$ with $T=0$.  At the finite region, point 
$L_1$ is 
the stable one.

Setting $\Omega_{DE}=1$, the system \eqref{syst1}-\eqref{syst2}  becomes a one-dimensional dynamical system: 
\begin{align}
   T' &=\frac{T^3 \sqrt{\frac{(T-1)^4}{T^4}-\frac{4 \beta ^2 c^2}{3}}}{1-T} \nonumber \\
   & -\frac{T^3 \sqrt{2 \beta ^2 (T-1)^4-\frac{8}{3} \beta ^4 c^2 T^4}}{(T-1)^2 \sqrt{3 (T-1)^2-\sqrt{9 (T-1)^4-12 \beta ^2 c^2
   T^4}}}.\label{1DDS}
\end{align}
The origin $T=0$ has eigenvalue $\lambda=1-\frac{1}{| c| }$. Moreover, the system admits,  at most, four additional equilibrium points $T_c$, with  $T_c\in\{T_1, T_2, T_3, T_4\}$ satisfying  
$\frac{(T-1)^4}{T^4}-\frac{4 \beta ^2 c^2}{3}=0$. Explicitly, we have that  
\begin{subequations}\label{valuesofT}
\begin{align}
T_{1,2}&= \frac{3}{3-4 \beta ^2 c^2}  -\frac{2 \sqrt{3} | c \beta | }{\left| 3-4 c^2 \beta ^2\right| } \nonumber \\
&  \mp \frac{\sqrt{2} \sqrt{12 | c \beta |  \left| 3-4 c^2 \beta ^2\right| +\sqrt{3} \left(16 \beta ^4
   c^4-9\right)} \sqrt{| c \beta | }}{\left| 3-4 c^2 \beta ^2\right| ^{3/2}}, \nonumber 
   \end{align}
\begin{align}
T_{3,4}&= \frac{3}{3-4 \beta ^2 c^2} + \frac{2 \sqrt{3} | c \beta | }{\left| 3-4 c^2 \beta ^2\right| } 
\\
&  \mp \frac{\sqrt{2} \sqrt{12 | c \beta |  \left| 3-4 c^2 \beta ^2\right| +\sqrt{3} \left(9-16
   \beta ^4 c^4\right)} \sqrt{| c \beta | }}{\left| 3-4 c^2 \beta ^2\right| ^{3/2}}. 
\end{align}
\end{subequations}
Such points with $0<T_c<1$, corresponding to de Sitter solution $a(t)\propto e^{H_0 t \left(\frac{1}{T_c}-1\right)}$, are stable for $c\geq 1$ and otherwise are saddle.  

\begin{figure} 
    \centering
    \includegraphics[width=8.5cm,scale=0.7]{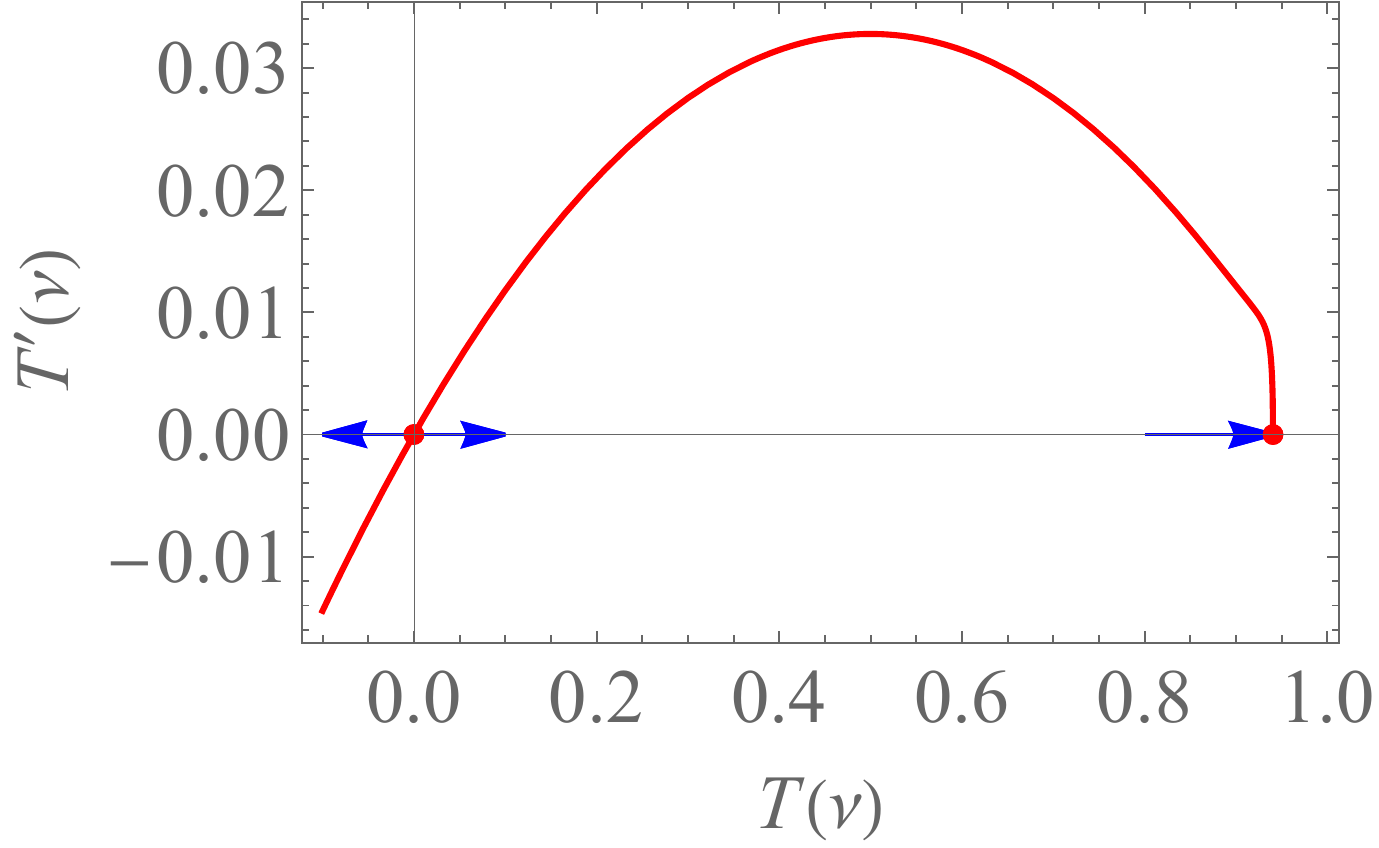}
    \caption{{\it{Phase-space plot of the one-dimensional dynamical system 
\eqref{1DDS}, for the best fit values $\beta = -0.003$ and $c = 1.151$ of Kaniadakis-holographic dark energy.  The  equilibrium point $T=0$ is unstable,
while the de Sitter equilibrium point 
$T=T_c\approx 0.941$ is stable.}}}
    \label{DS1D}
\end{figure}

For the 
best-fit values $\beta = -0.003$ and $c = 1.151$, the origin has eigenvalue $ \lambda \approx 0.13$, and therefore it is a source. In this case the only real value is $T_3 \approx 0.941$. The exact eigenvalue is negative infinity (for $c\geq 1$) at the exact value of 
$T_3$,  and therefore it is stable.
In Fig. \ref{DS1D}  we draw a phase-space plot of the one-dimensional dynamical system \eqref{1DDS}, for the best fit values $\beta = -0.003$ and $c = 1.151$ of Kaniadakis holographic dark energy. The   equilibrium point $T=0$ is unstable,
while the de Sitter equilibrium point 
$T=T_c\approx 0.941$ is stable.

%%%%%%%%%%%%%%%%%%%%%%%%%%%%%%%%%%%%%%%%%
\subsection{Global dynamical systems formulation}

In the previous subsection we performed the local analysis 
of the scenario. However, due to the presence of  rational functions 
  that are not analytic in the whole domain, it becomes necessary to investigate 
the full  global dynamics.
We start by defining the dimensionless variables $\theta, T$ as  
\begin{align}
T= \frac{H_0}{H+ H_0}= \frac{1}{1+E}, \quad     \theta= \arcsin \left(\sqrt{1- 
\frac{\rho_{DE}}{3 M_p^2 H^2 }}\right),
\end{align}
such that 
\begin{equation}
     \sin^2 (\theta)= \frac{\rho_m}{3 M_p^2 H^2 }, \quad   \cos^2 (\theta)= \frac{\rho_{DE}}{3 M_p^2 H^2 }.
\end{equation}
For an expanding universe ($H>0$), we have that $T\in[0,1]$, while  $\theta$ 
is a periodic coordinate and, thus, we can set $\theta\in[-\pi, \pi]$. 
Therefore, we obtain a global phase-space formulation.

%%%%%%%%%%%%%%%%%%%%%%%%%%%%%%%%%%%%%%%%%
\subsubsection{Standard holographic dark energy ($\beta=0$)}
\label{sect4.3.1}

In order to present the features of Kaniadakis-holographic dark energy in 
comparison with   standard-holographic dark energy, we first analyze the latter 
case for completeness, namely  we consider the  system 
\eqref{ODEK0}-\eqref{HK0}   for $\beta=0$. In this case,  we obtain 
\begin{align}
& T^{\prime}=\frac{(T-1) T \left\{\cos ^2(\theta ) [(3 w_m+1)c +2 \cos (\theta 
)]-3 c (w_m+1)\right\}}{2 c}, \label{Case1_a}\\
& \theta^{\prime}=-\frac{[(3 w_m+1)c+2 \cos (\theta )] \sin (2 \theta )}{4 c}.  
\label{Case1_b} 
\end{align}

The critical points of the above system, alongside their associated 
eigenvalues, are presented in Table \ref{tab:my_label1}. 
 Note that $\theta$ is unique modulo $2\pi$, and  focus on 
$\cos \theta\geq 0$. In the following list	$\arctan[x,y]$
gives the arc tangent of $y/x$, taking into account on which quadrant the point $(x,y)$ is in.  When $x^2+y^2=1$, $\arctan[x,y]$ gives the number $\theta$  such that $x=\cos\theta$ and $y=\sin\theta$.
\begin{table*}
       \centering
       \caption{ \label{tab:my_label1}
     The critical  points and their associated eigenvalues of the system 
\eqref{Case1_a}-\eqref{Case1_b}  for $\beta=0$ in 
\eqref{ODEK0}-\eqref{HK0}, namely for the case of standard holographic dark 
energy. We use the notation $x=\frac{1}{2}c (3 w_m+1)$, while $c_1\in \mathbb{Z}$.}
    \begin{tabular}{|c|c|c|}\hline
   Label&   $ (T,\theta)$ & Eigenvalues\\\hline
     $P_1$ &     $\left( 0, 2 \pi  c_1\right)$ & $\left\{\frac{c-1}{c},-\frac{3
   w_m c+c+2}{2 c}\right\}$ \\ 
   $P_2$ &  $\left( 0, \frac{1}{2} \pi  \left(4 c_1-1\right)\right)$ &
   $\left\{\frac{3 (w_m+1)}{2},\frac{1}{2} (3 w_m+1)\right\}$ \\ 
  $P_3$ &   $\left(0, \frac{1}{2} \pi  \left(4 c_1+1\right)\right)$ &
   $\left\{\frac{3 (w_m+1)}{2},\frac{1}{2} (3 w_m+1)\right\}$ \\ 
  $P_4^\pm$ &  $ \left(0,  2 \pi  c_1 \pm \pi \right)$ & $\left\{1+\frac{1}{c},-\frac{3
   w_m}{2}+\frac{1}{c}-\frac{1}{2}\right\} $\\ 
  $P_5$ &   $\left(0,  \arctan\left[-x,-\sqrt{1-x^2}\right]+2 \pi  c_1\right)$  & $\left\{\frac{3 (w_m+1)}{2},\frac{1}{8}
   (3 w_m+1) \left(c^2(1+3 w_m)^2-4\right)\right\}$ \\ 
   $P_6$ &  $ \left(0,  \arctan\left[-x,\sqrt{1-x^2}\right]+2 \pi  c_1\right)$ & $\left\{\frac{3 (w_m+1)}{2},\frac{1}{8}
   (3 w_m+1) \left(c^2(1+3 w_m)^2-4\right)\right\}$ \\ 
  $P_7$ &  $ \left(1, 2 \pi  c_1\right)$ & $\left\{\frac{1}{c}-1,-\frac{3
   w_m c+c+2}{2 c}\right\}$ \\ 
  $P_8$ &  $\left(1,  \frac{1}{2} \pi  \left(4 c_1-1\right)\right)$ &
   $\left\{-\frac{3}{2} (w_m+1),\frac{1}{2} (3 w_m+1)\right\}$ \\ 
  $P_9$ &   $\left(1, \frac{1}{2} \pi  \left(4 c_1+1\right)\right)$ &
   $\left\{-\frac{3}{2} (w_m+1),\frac{1}{2} (3 w_m+1)\right\}$ \\ 
  $P_{10}^\pm$ &  $ \left(1,  2 \pi  c_1\pm\pi \right)$ & $\left\{-\frac{c+1}{c},-\frac{3
   w_m}{2}+\frac{1}{c}-\frac{1}{2}\right\}$ \\ 
   $P_{11}$ & $ \left(1,  \arctan\left[-x,-\sqrt{1-x^2}\right]+2 \pi  c_1\right)$ & $\left\{-\frac{3}{2} (w_m+1),\frac{1}{8}
   (3 w_m+1) \left(c^2(1+3 w_m)^2-4\right)\right\}$ \\ 
  $P_{12}$ & $ \left(1, \arctan\left[-x,\sqrt{1-x^2}\right]+2 \pi  c_1\right) $& $\left\{-\frac{3}{2} (w_m+1),\frac{1}{8}
   (3 w_m+1) \left(c^2(1+3 w_m)^2-4\right)\right\}$ \\\hline
    \end{tabular}
\end{table*}

In summary, in the case $\beta=0$,  
the critical points can be completely characterized. In particular:
 
 \begin{itemize}
     \item  Point $P_1$  always exists. It corresponds to a dark-energy 
dominated  solution, i.e. $\Omega_{DE}=1$ with $T=0$. It is a stable point for 
$-1<w_m<1, \quad 0<c<1$. 
     
     \item Points $P_2$ and $P_3$ exist always. They are two representations 
of the matter-dominated solution $\Omega_{DE}=0$ with $T=0$. They are past 
attractors, i.e. unstable points, for $-\frac{1}{3}<w_m\leq 1$, while they are 
saddle for $-1<w_m<-\frac{1}{3}$. 
     
  \item Points $P_4^\pm$ exist always. They correspond to the dark-energy 
dominated  solution with $\Omega_{DE}=1$ with $T=0$. They are unstable points 
for $0<c<\frac{1}{2},  -1\leq w_m\leq
   1$, or $ c\geq \frac{1}{2}, \quad -1\leq w_m<\frac{2-c}{3 c}$, while they are 
saddle for $c>\frac{1}{2}, \quad \frac{2-c}{3 c}<w_m\leq
   1$. 
     
     \item Points $P_5$ and $P_6$  exist for  $-1\leq \frac{1}{2}c (3 w_m+1)\leq 1$.  They are sources for $0\leq c\leq 1, \quad  -1<w_m<-\frac{1}{3}$ or $c>1, \quad  -\frac{c+2}{3
   c}<w_m<-\frac{1}{3}$. 
For $0\leq c<\frac{1}{2},\quad -\frac{1}{3}<w_m\leq 1$, or $c\geq \frac{1}{2}, \quad
   -\frac{1}{3}<w_m<\frac{2-c}{3 c}$,  they are saddle.

     \item Point $P_7$ exists always. It corresponds to a dark-energy  dominated 
 solution $\Omega_{DE}=1$ with $T=1$. It is a stable point for $c>1,  \quad -\frac{c+2}{3 c}<w_m\leq 1$.
     
     \item Points $P_8$ and $P_9$ exist always. They are two representations 
of the matter-dominated solution $\Omega_{DE}=0$ with $T=1$. They are stable 
points for $-1<w_m<-\frac{1}{3}$, while they are saddle points for $-\frac{1}{3}<w_m\leq 1$. 
     
     \item Points $P_{10}^\pm$ are two representations  of the matter-dominated 
solution $\Omega_{DE}=0$ with $T=1$. They are stable points for $c>\frac{1}{2}, 
\frac{2-c}{3 c}<w_m\leq 1$, while they are saddle for $0<c<\frac{1}{2},  \quad 
-1\leq w_m\leq 1$, or $c\geq \frac{1}{2}, \quad -1\leq w_m<\frac{2-c}{3
   c}$. 
     
     \item Points $P_{11}$ and $P_{12}$ exist for
     $-1\leq \frac{1}{2}c (3 w_m+1)\leq 1$.
     They are saddle for $0\leq c\leq 1, \quad  -1<w_m<-\frac{1}{3}$ or $c>1, \quad  -\frac{c+2}{3
   c}<w_m<-\frac{1}{3}$, while for $0\leq c<\frac{1}{2},\quad -\frac{1}{3}<w_m\leq 1$, or $c\geq \frac{1}{2}, \quad
   -\frac{1}{3}<w_m<\frac{2-c}{3 c}$,  they are stable. 
          
 \end{itemize}

    \begin{figure}
    \centering
    \includegraphics[width=8cm,scale=0.6]{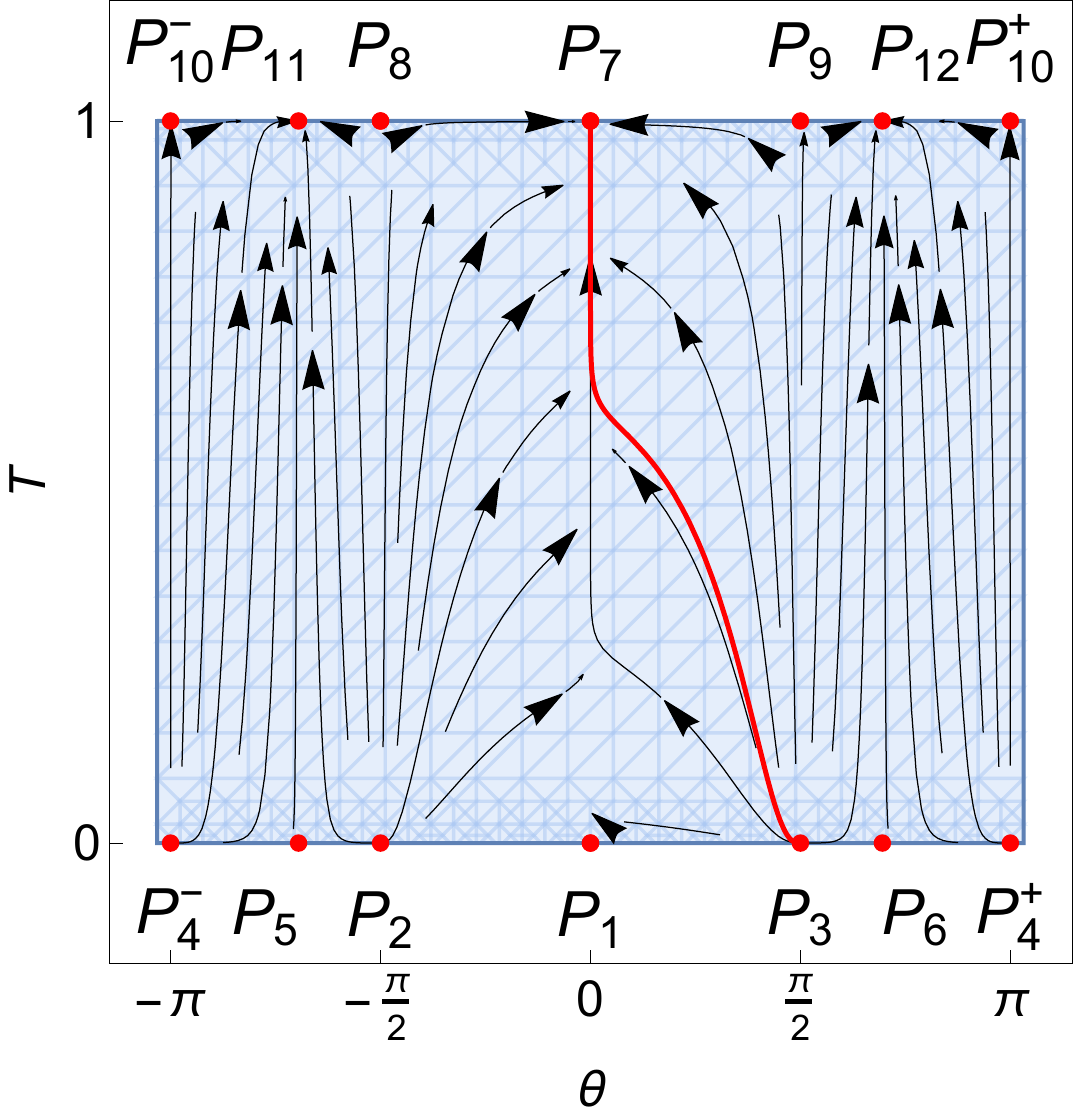}
    \caption{ {\it{Phase-space plot of the dynamical system 
\eqref{Case1_a}-\eqref{Case1_b}  for 
$\beta=0$ in 
\eqref{ODEK0}-\eqref{HK0}, namely for standard holographic dark energy,  for the  value $c = 1.151$, and for dust 
matter $w_m = 0$. The red curve represents the solution for the initial data 
$\Omega_{DE}|_{z=0}=0.71$ (i.e., $\theta 
(0)=\arccos\left(\frac{1}{10}\sqrt{71}\right)\approx 0.569$), corresponding to the mean value 
obtained with 
the joint analysis CC+SNIa+BAO,  and for $T|_{z=0}=0.5$. The 
dashed blue region is the physical region  where the equations are real-valued. 
}}}
    \label{DS3}
\end{figure}
 
In order to give a better picture  of the system behavior, 
Fig. \ref{DS3}
display  a phase-space plot of the system \eqref{Case1_a}-\eqref{Case1_b}  for 
$\beta=0$ in 
\eqref{ODEK0}-\eqref{HK0}, 
and 
dust matter. The red curve  
corresponds to the universe evolution according to parameter mean values from the joint analysis. 
From this figure we deduce that   the late-time attractor is the dark-energy dominated solution  with $\Omega_{DE}=1$ and $T=1$ (point $P_7$), while the past attractor 
is the matter-dominated solution with $\Omega_{DE}=0$ and $T=0$ (point $P_3$).
For other initial conditions there are other late-time attractors, such as points $P_{11}$ and $P_{12}$ 
which are stable 
for the best-fit parameters 
since they satisfy $c\geq \frac{1}{2}, \quad
   -\frac{1}{3}<w_m<\frac{2-c}{3 c}$. 
   These points   are scaling solutions since they have
   $\Omega_{DE}= x^2$ and $\Omega_{DM}= 1-x^2$, with $x=\frac{c}{2} (3 w_m+1)= \frac{c}{2} $ for $w_m=0$.  
   Additionally, points $P_2$, $P_3$, which are matter-dominated solutions, and points $P_{4}^\pm$, which are dark-energy dominated solutions,  are also  past attractors. %too. 

%%%%%%%%%%%%%%%%%%%%%%%%%%%%%%%%%%%%%%%%%
\subsubsection{Kaniadakis holographic dark energy ($\beta\neq 0$)}

Let us now investigate the full extended model of Kaniadakis holographic 
dark energy, namely the general case where $\beta\neq 0$.
The full system \eqref{eq2.15}-\eqref{eq2.16} becomes 
\begin{eqnarray}
T^{\prime}= & \frac{3}{2} (1-T) T (w_m+1) \sin ^2(\theta ) +\frac{T^3 \sqrt{\frac{(1-T)^4 \cos
   ^4(\theta )}{T^4}-\frac{4 \beta ^2 c^2}{3}}}{1-T}\nonumber \\
& -\frac{  T^5 \sqrt{\frac{18 (1-T)^4 \cos ^4(\theta )\beta^2}{T^4}-24 \beta ^4 c^2}}{3 (T-1)^2 \sqrt{3 (T-1)^2 \cos ^2(\theta )-\sqrt{9 (1-T)^4 \cos ^4(\theta )-12 \beta ^2 c^2 T^4}}}, \label{FullT}
\end{eqnarray}
\begin{eqnarray}
\theta^{\prime}=& -\frac{3}{4} (w_m+1) \sin (2 \theta )+\frac{T^2 \tan (\theta ) \sqrt{\frac{(1-T)^4 \cos ^4(\theta )}{T^4}-\frac{4 \beta ^2
   c^2}{3}}}{(T-1)^2} \nonumber \\
& -\frac{\sqrt{\frac{2}{3}} T^2 \tan (\theta ) \sqrt{-\beta ^2 \left(4 \beta ^2 c^2 T^4-3 (1-T)^4 \cos ^4(\theta )\right)}}{(1-T)^3 \sqrt{3 (T-1)^2 \cos ^2(\theta )-\sqrt{9 (1-T)^4 \cos ^4(\theta )-12 \beta ^2
   c^2 T^4}}}. \label{Fulltheta}
\end{eqnarray}
Moreover, the physical region of the phase space is 
\begin{equation}
    3 (1-T)^4 \cos ^4(\theta )-4 \beta ^2
   c^2 T^4\geq 0. \label{region}
\end{equation}

We proceed by  studying the critical points of the system 
\eqref{FullT}-\eqref{Fulltheta}  in the physical region  
\eqref{region}  and their stability. We mention that for $\beta \neq 0$ the 
invariant set 
$T=1$ is not physical. 
Near the invariant set $T=0$ the system \eqref{FullT}-\eqref{Fulltheta}  becomes
\begin{align}
  & T'=\left[-\frac{\cos ^3(\theta )}{c}+\cos ^2(\theta )+\frac{3}{2} (w_m+1)
   \sin ^2(\theta )\right] T+O\left(T^2\right),\\
   & \theta'=-\frac{[(3 w_m+1)c+2 \cos
   (\theta )] \sin (2 \theta )}{4 c}+O\left(T^2\right).
\end{align}
In Table \ref{tab:my_label2} we summarize the critical points $P_1$ to $P_6$, alongside their 
associated eigenvalues. Furthermore, the stability conditions are the same as 
discussed in subsection \ref{sect4.3.1}.
In summary, in the invariant set $T=0$,  the critical points are:
 
 \begin{itemize}
     \item  Point $P_1$  exists always. It corresponds to a dark-energy 
dominated  solution, i.e. $\Omega_{DE}=1$ with $T=0$. It is a stable point for 
$-1<w_m<1, \quad 0<c<1$. 
     
     \item Points $P_2$ and $P_3$ exist always. They are two representations 
of the matter-dominated solution $\Omega_{DE}=0$ with $T=0$. They are past 
attractors, i.e. unstable points, for $-\frac{1}{3}<w_m\leq 1$, while they are 
saddle for $-1<w_m<-\frac{1}{3}$. 
     
  \item Points $P_4^\pm$ exist always. They correspond to the dark-energy 
dominated  solution with $\Omega_{DE}=1$ with $T=0$. They are unstable points 
for $0<c<\frac{1}{2}, \quad -1\leq w_m\leq
   1$, or $ c\geq \frac{1}{2}, \quad -1\leq w_m<\frac{2-c}{3 c}$, while they are 
saddle for $c>\frac{1}{2}, \frac{2-c}{3 c}<w_m\leq
   1$.

     \item Points $P_5$ and $P_6$  exist for  $-1\leq \frac{1}{2}c (3 w_m+1)\leq 1$.  They are unstable 
     for $0\leq c\leq 1, \quad  -1<w_m<-\frac{1}{3}$ or $c>1, \quad  -\frac{c+2}{3
   c}<w_m<-\frac{1}{3}$, while for $0\leq c<\frac{1}{2},\quad -\frac{1}{3}<w_m\leq 1$, or $c\geq \frac{1}{2}, \quad
   -\frac{1}{3}<w_m<\frac{2-c}{3 c}$,  they are saddle. 
     
\end{itemize}
Moreover, the system admits,  at most, twelve additional equilibrium points $(\theta, T)$, with $\theta\in 
\{\theta_1, \theta_2, \theta_3\}$  satisfying 
$\cos^2(\theta)=1$, and $T\in\{T_1, T_2, T_3, T_4\}$ satisfying  
$\frac{(T-1)^4}{T^4}-\frac{4 \beta ^2 c^2}{3}=0$, explicitly  given by \eqref{valuesofT}. 
Such points with $0<T_c<1$, corresponding to de Sitter solution $a(t)\propto e^{H_0 t \left(\frac{1}{T_c}-1\right)}$, are stable for $c\geq 1$ or saddle otherwise.

Notice that the physical values are the real values of $T_i$ satisfying 
$0\leq T_i\leq 1$, $i=1,2,3,4$. One eigenvalue is always $-\frac{3}{2} (1 + 
w_m)$, while the other one is infinite.  The stability conditions are found 
numerically and, moreover, for $\beta=0$ we find $T_i=0$.
Hence, we re-obtain points $P_7$ and $P_{10}^{\pm}$ in Table \ref{tab:my_label1}. Indeed, for $\beta=0$ all the results of section \ref{sect4.3.1}  are recovered. 
\begin{table*}
      \centering
    \begin{tabular}{|c|c|c|}\hline
   Label&   $ (T,\theta)$ & Eigenvalues\\\hline
     $P_1$ &     $\left( 0, 2 \pi  c_1\right)$ & $\left\{\frac{c-1}{c},-\frac{3
   w_m c+c+2}{2 c}\right\}$ \\ 
   $P_2$ &  $\left( 0, \frac{1}{2} \pi  \left(4 c_1-1\right)\right)$ &
   $\left\{\frac{3 (w_m+1)}{2},\frac{1}{2} (3 w_m+1)\right\}$ \\ 
  $P_3$ &   $\left(0, \frac{1}{2} \pi  \left(4 c_1+1\right)\right)$ &
   $\left\{\frac{3 (w_m+1)}{2},\frac{1}{2} (3 w_m+1)\right\}$ \\ 
  $P_4^\pm$ &  $ \left(0,  2 \pi  c_1 \pm \pi \right)$ & $\left\{1+\frac{1}{c},-\frac{3
   w_m}{2}+\frac{1}{c}-\frac{1}{2}\right\} $\\ 
  $P_5$ &   $\left(0, 
   \arctan\left[-x,-\sqrt{1-x^2}\right]+2 \pi  c_1\right)$  & $\left\{\frac{3 
(w_m+1)}{2},\frac{1}{8}
   (3 w_m+1) \left(c^2(1+3 w_m)^2-4\right)\right\}$ \\ 
   $P_6$ &  $ \left(0,  
    \arctan
    \left[-x,\sqrt{1-x^2}\right]+2 \pi  c_1\right)$ & $\left\{\frac{3 
(w_m+1)}{2},\frac{1}{8}
   (3 w_m+1) \left(c^2(1+3 w_m)^2-4\right)\right\}$  \\\hline
    \end{tabular}
      \caption{ \label{tab:my_label2}
       The critical  points and their associated eigenvalues of the system 
\eqref{FullT}-\eqref{Fulltheta} in the invariant set $T=0$.   
We use the notation $x=\frac{1}{2}c (3 w_m+1)$, $c_1\in \mathbb{Z}$.}
\end{table*}

The solutions of physical interest are those with 
$T=0$. Point $P_1$, which corresponds to a dark-energy dominated  solution 
$\Omega_{DE}=1$ with $T=0$, is stable for $-1<w_m<1, \quad 0<c<1$. Points $P_2$ 
and $P_3$, which are two representations of the matter-dominated solution 
$\Omega_{DE}=0$ with $T=0$, are past attractors for $-\frac{1}{3}<w_m\leq 1$ or 
saddle for $-1<w_m<-\frac{1}{3}$.
 Points $P_4^\pm$, which correspond to a dark-energy dominated  solution 
 are unstable for $0<c<\frac{1}{2}, \quad -1\leq w_m\leq
   1$, or $ c\geq \frac{1}{2}, \quad -1\leq w_m<\frac{2-c}{3 c}$, while they are 
saddle points for $c>\frac{1}{2}, \quad \frac{2-c}{3 c}<w_m\leq
   1$. Finally, points $P_5$ and $P_6$ exist for  $-1\leq \frac{1}{2}c (3 w_m+1)\leq 1$.  They are sources for $0\leq c\leq 1, \quad  -1<w_m<-\frac{1}{3}$ or $c>1, \quad  -\frac{c+2}{3
   c}<w_m<-\frac{1}{3}$, while for $0\leq c<\frac{1}{2},\quad -\frac{1}{3}<w_m\leq 1$, or $c\geq \frac{1}{2}, \quad
   -\frac{1}{3}<w_m<\frac{2-c}{3 c}$, they are saddle. 
Finally, note that the region where $T\rightarrow 1$ is contained in the 
complex-valued domain.  This forbids solutions with $H=0$, which appear in the 
standard-holographic dark energy scenario of \eqref{Case1_a}-\eqref{Case1_b}.

\begin{figure}
    \centering
    \includegraphics[width=8cm,scale=0.6]{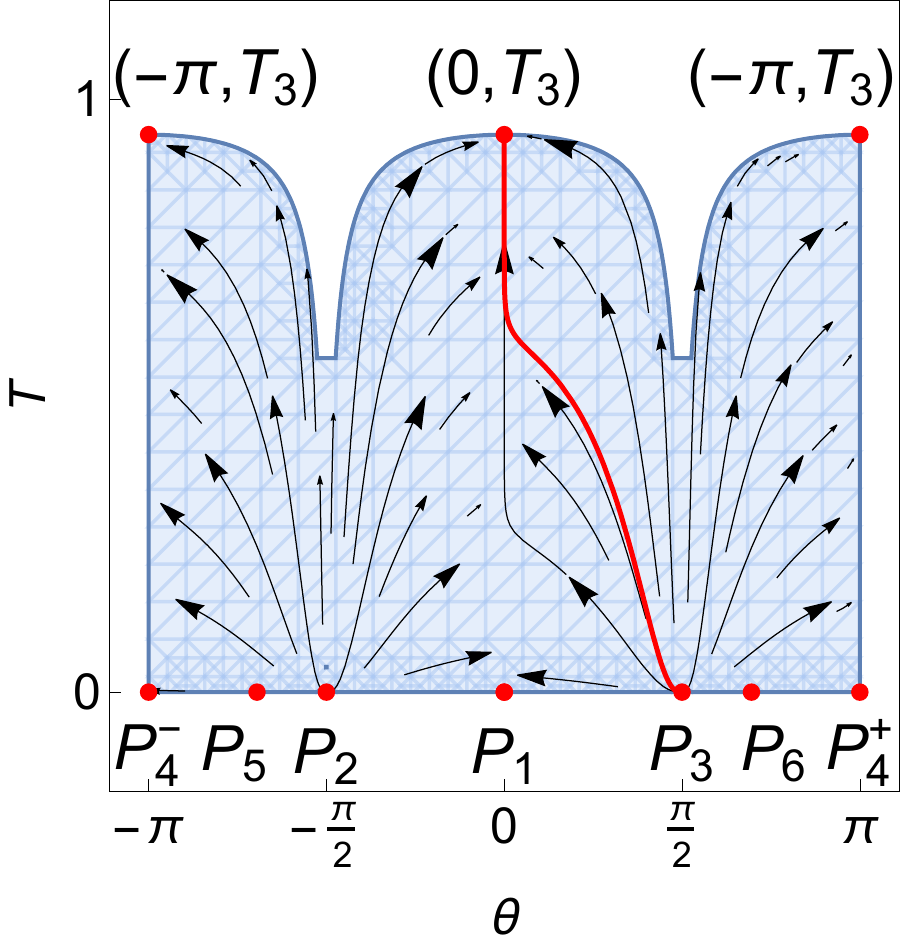}
    \caption{{\it{Phase-space plot of the dynamical system 
\eqref{FullT}-\eqref{Fulltheta}  
for the best fit values 
$\beta =-0.003$ and $c = 1.151$, and for dust matter ($w_m = 0$).
The red curve represents the solution for the initial data 
$\Omega_{DE}|_{z=0}=0.71$ (i.e., $\theta 
(0)=\arccos\left(\frac{1}{10}\sqrt{71}\right)\approx 0.569$), corresponding to the mean value from the joint analysis CC+SNIa+BAO,  and for $T|_{z=0}=0.5$.  The 
dashed-blue region is the physical region  where the equations are real-valued.  
}}}
\label{DS2}
\end{figure}
In Fig. \ref{DS2} we 
show  a phase-space plot of the system  \eqref{FullT}-\eqref{Fulltheta} for the 
best-fit values $\beta = -0.003$ 
and $c = 1.151$ and for dust matter ($w_m = 0$). 
In this case the only real value is $T_3 \approx 0.941$.  At points $(-\pi, 
T_3)$, $(0, T_3)$, and $(\pi, T_3)$, the eigenvalues are $-\frac{3}{2}$ and one eigenvalue is negative infinity at the exact value of 
$T_3$,   therefore they are sink. For comparison, we have added
the red curve, corresponding to the solution for the initial data 
$\Omega_{DE}|_{z=0}=0.71$ (i.e., $\theta 
(0)=\arccos\left(\frac{1}{10}\sqrt{71}\right)\approx 0.569$), which is the mean value from the joint analysis CC+SNIa+BAO,  and for $T|_{z=0}=0.5$. From this figure it is confirmed that the late-time attractor is the 
dark-energy dominated solution (de Sitter solution with $a(t)\propto e^{H_0 t \left(\frac{1}{T_c}-1\right)}, H_0= h \times 100\,\mathrm{km\, s}^{-1} \mathrm{Mpc}^{-1}, T_c\approx 0.941, h= 0.761$), while the past 
attractor 
is the matter-dominated solution.

%%%%%%%%%%%%%%%%%%%%%%%%%%%%%%%%%%%%%%%%
\section{Summary and discussion} \label{sec:Con}
%%%%%%%%%%%%%%%%%%%%%%%%%%%%%%%%%%%%%%%%%

We investigated the scenario of Kaniadakis-holographic dark energy scenario by 
confronting it with  observational data. 
This is an extension of the usual holographic dark-energy model which 
arises 
from 
the use of the generalized Kaniadakis entropy instead of the 
standard  Boltzmann-Gibbs  one, which in turn 
appear 
from the relativistic 
extension of standard statistical 
theory. 

We applied the Bayesian approach
to extract the likelihood bounds
of the Kaniadakis parameter, as well as 
the other free model parameters. In 
particular, we performed a Markov Chain Monte Carlo analysis using 
 data from cosmic chronometers,  supernovae  type Ia, and  
Baryon Acoustic Oscillations observations.
Concerning the Kaniadakis parameter, we 
found 
that it  is constrained 
around 0, namely, around the value in which Kaniadakis entropy 
recovers the standard Bekenstein-Hawking one, as expected. Additionally, for  
$\Omega_m^{(0)}$ we obtained  a slightly smaller value 
compared 
to $\Lambda$CDM scenario. 

Furthermore, we reconstructed the evolution of the Hubble, 
deceleration and jerk parameters in the redshift range $0<z<2$. 
We find that,
within one 
sigma confidence level with those reported in \citet{Herrera-Zamorano:2020}, the deceleration-acceleration transition redshift is 
$z_T = 0.86^{+0.21}_{-0.14}$, 
and 
the age of the  Universe
is $t_U = 13.000^{+0.406}_{-0.350}\,\rm{Gyrs}$.
Lastly, we  applied the usual information criteria in order to compare the 
statistical significance of the fittings with $\Lambda$CDM cosmology. Both criteria AICc and BIC conclude that the $\Lambda$CDM scenario is strongly favored
in comparison 
to Kaniadakis-holographic dark energy.

Finally, we performed a detailed dynamical-system analysis 
to extract 
the local and global features of the evolution in the scenario of Kaniadakis-holographic dark energy. We extracted the critical points as well as their 
stability properties and found 
that  
the   past 
attractor of the Universe is the matter-dominated solution, while the late-time 
stable solution is the  dark-energy-dominated one with $H\rightarrow 0$.

In summary, Kaniadakis-holographic dark energy presents interesting 
cosmological behavior and is in agreement with observations. We remark that the scenario may solve the turning point in the Hubble parameter reconstruction of   standard holographic dark energy \citep{Colgain:2021beg}, which violates the NEC, and thus it is an interesting improvement in this context.

\section*{Acknowledgements}
\addcontentsline{toc}{section}{Acknowledgements}

We thank the anonymous referee for thoughtful remarks and suggestions. Authors acknowledge Eoin O. Colgain for fruitful comments. G.L. was funded by  Agencia Nacional de Investigaci\'on y Desarrollo - ANID for financial support through the program FONDECYT Iniciaci\'on grant no. 11180126 and by Vicerrectoría de Investigación y Desarrollo Tecnológico at UCN. J.M. acknowledges the support from ANID project Basal AFB-170002 and ANID REDES 190147. M.A.G.-A. acknowledges support from SNI-M\'exico, CONACyT research fellow, ANID REDES (190147), C\'atedra Marcos Moshinsky and Instituto Avanzado de Cosmolog\'ia (IAC). A.H.A. thanks to the PRODEP project, Mexico for resources and financial support and thanks also to the support from Luis Aguilar, Alejandro de Le\'on, Carlos Flores, and Jair Garc\'ia of the Laboratorio Nacional de Visualizaci\'on Cient\'ifica Avanzada. V.M. acknowledges support from Centro de Astrof\'{\i}sica de Valpara\'{i}so and ANID REDES 190147. This work is partially supported by the Ministry of Education and Science of the Republic of Kazakhstan, Grant AP08856912.

%%%%%%%%%%%%%%%%%%%%%%%%%%%%%%%%%%%%%%%%%%%%%%%%%%

\section*{Data Availability}
The data underlying this article were cited in Section \ref{sec:Data}.



\bibliographystyle{mnras}
\bibliography{main} % if your bibtex file is called example.bib

% Don't change these lines
\bsp	% typesetting comment
\label{lastpage}
\end{document}